\documentclass[showpacs,preprintnumbers,amsmath,aps,nofootinbib,amssymb,superscriptaddress]{revtex4}
\usepackage{graphicx}
\usepackage[english]{babel}
\usepackage{bm}
\usepackage[font=scriptsize]{caption}
\usepackage{ragged2e}
\usepackage[font=small,labelfont=bf,justification=raggedright]{caption}
\usepackage[utf8]{inputenc}
\usepackage[tikz]{bclogo}
\usepackage[framemethod=tikz]{mdframed}
\usepackage{booktabs}
\usepackage{array}
\usepackage{amsmath}
\usepackage{tikz}

\newsavebox{\measurebox}
\begin{document}
\baselineskip 16pt
\title{ Power law scalar potential in the Saez-Ballester like theory: Exact solutions in the Bianchi type I case}
\author{J. Socorro}
\email{socorro@fisica.ugto.mx}
 \affiliation{Departamento de
F\'{\i}sica, DCeI, Universidad de Guanajuato-Campus Le\'on, C.P.
37150, Le\'on, Guanajuato, M\'exico}

\author{A. Gil-Ocaranza}
\email{victor.gil@academicos.udg.mx} \affiliation{Departamento de Física, Centro de Investigación y de Estudios Avanzados del Instituto
Politécnico Nacional
Apartado Postal 14-740, 07000, Ciudad de México, México.}
\affiliation{Centro Universitario de la Costa - UDG
Av. Universidad de Guadalajara 203, Ixtapa, Los Tamarindos, 48280 Puerto Vallarta, Jal.,México.}
\author{Juan Luis P\'erez $^1$ }
\email{jl.perezp@ugto.mx}

\begin{abstract}

We investigate exact anisotropic Bianchi type I cosmological solutions in a generalized S'aez--Ballester--K-essence-like theory containing two interacting scalar fields with quintessence, phantom, and mixed (quintom) kinetic sectors. The scalar-field self-interaction is described by inverse power-law potentials, which admit analytical solutions after an appropriate transformation of the field equations. The mixed kinetic interaction imposes nontrivial constraints on the model parameters, leading to six exact cosmological branches associated with the quintessence, phantom, and quintom scenarios.

The exact solutions provide explicit expressions for the effective volume and scalar fields, allowing the reconstruction of the Hubble parameter, the deceleration parameter, and the effective equation-of-state parameter. Although the six branches exhibit distinct transient behaviors, all physically admissible solutions evolve toward the same asymptotic de Sitter attractor characterized by accelerated expansion. The analytical solutions reveal how the different kinetic sectors control the cosmological evolution while preserving a common late-time behavior. These results provide a unified analytical description of anisotropic cosmological dynamics in generalized scalar-field theories and may serve as a useful framework for investigating the role of interacting scalar fields in the early and late evolution of the Universe.

\end{abstract}

\maketitle

\section{Introduction}

Scalar fields constitute one of the most important ingredients of
modern theoretical cosmology. They provide a unified framework for
describing the inflationary stage of the early universe, the present
accelerated expansion, and a wide class of modified theories of
gravity. Depending on the structure of their kinetic sector and
self-interaction potential, scalar-field models naturally give rise to
quintessence, phantom, and quintom cosmologies, each exhibiting
distinct dynamical properties. Although numerous numerical analyses and
dynamical-system studies have been devoted to these scenarios
\cite{CQG-2014,PRD-2017,PDU-2024}, exact analytical solutions remain
comparatively scarce, particularly for anisotropic cosmological models
containing more than one interacting scalar field.

Among the various scalar-field theories proposed in the literature,
K-essence models constitute one of the most general noncanonical
extensions of Einstein gravity. Their dynamics are described by the
action \cite{IJTP-2014,roland,chiba,bose,arroja,tejeiro}
\begin{equation}
S_K=\int d^4x\,\sqrt{-g}\,
\left[f(\phi)\,{\cal G}(X)\right],
\end{equation}
where
\begin{equation}
{\cal G}(X)=X
=-\frac12g^{\mu\nu}\nabla_\mu\phi\nabla_\nu\phi,
\end{equation}
and $f(\phi)$ is an arbitrary function of the scalar field. Depending
on the choice of $f(\phi)$, this framework admits a rich variety of
exact classical solutions \cite{AHEP-2014}, quantum cosmological
models \cite{IJTP-2014}, fractional extensions
\cite{universe-2023,Fractal-2023}, and noncommutative generalizations
\cite{universe-2024,GRG-2025}.

A particular realization of the K-essence formalism is provided by the
Saez--Ballester theory \cite{saez-ballester-1986}, whose action can be
written as
\begin{equation}
S_{SB}
=\int d^4x\,\sqrt{-g}\,
\left[\phi^{m}X\right],
\end{equation}
with $m\in\mathbb{R}$. This model has received considerable attention
owing to the existence of exact classical and quantum solutions for
Friedmann--Robertson--Walker and Bianchi Class A cosmologies
\cite{fizika-2010}, as well as anisotropic Bianchi models
\cite{rmf-2010}. More recently, multifield extensions have attracted
increasing interest because the interaction between scalar degrees of
freedom enriches the cosmological dynamics and naturally leads to
quintessence, phantom, and quintom scenarios. A general chiral
multifield action takes the form
\begin{equation}
S=\int d^4x\,\sqrt{-g}\,
\left[
-\frac12\nabla_\mu\phi\nabla^\mu\phi
-\frac12g^{\mu\nu}F(\phi)
\nabla_\mu\psi\nabla_\nu\psi
-V(\phi)
\right],
\end{equation}
where the coupling function $F(\phi)$ is commonly chosen as an
exponential \cite{GC-2012,QM-2013}, a hyperbolic function
\cite{coupled}, or more general families of functions
\cite{PRD-2004-chimento}.

Despite these developments, exact anisotropic solutions involving
chiral multifield cosmologies with negative power-law scalar potentials
remain largely unexplored. Most existing works either employ
exponential potentials or investigate the dynamics through phase-space
methods, whereas exact analytical solutions for negative power-law
potentials in anisotropic multifield cosmology are still absent from
the literature.

The purpose of the present work is to fill this gap by combining the
Saez--Ballester and K-essence formalisms within a chiral multifield
framework. We consider two interacting scalar fields endowed with
negative power-law self-interaction potentials,
\begin{equation}
V(\psi_1,\psi_2)
=
V_1\psi_1^{-\lambda_1}
+
V_2\psi_2^{-\lambda_2},
\end{equation}
and study their cosmological evolution in an anisotropic Bianchi type I
spacetime. Through the transformation $\psi=e^\phi$, the model can be
recast into a generalized Saez--Ballester--K-essence-like theory, whose
action is
\begin{equation}
S_{SBK}
=
\int d^4x\,\sqrt{-g}\,
\left[
{\cal G}({\cal X})
+
V_0\psi^{\pm\lambda}
\right],
\label{SBK}
\end{equation}
where
\begin{equation}
{\cal X}
=
\epsilon\psi^{-2}X
=
-\frac12\epsilon\psi^{-2}
g^{\mu\nu}
\nabla_\mu\psi
\nabla_\nu\psi,
\end{equation}
with $\epsilon=+1$ describing quintessence fields and $\epsilon=-1$
corresponding to phantom fields. Under the transformation
$\psi=e^\phi$, the kinetic sector reduces to the usual canonical or
phantom form, while the power-law potential is mapped into an
exponential potential. This relation establishes a direct connection
between the present formalism and previous exact solutions obtained for
exponential scalar-field cosmologies \cite{CQG-2021,IJMPD-2021}.

Unlike previous analyses, the present approach yields exact analytical
solutions for the quintessence, phantom, and quintom sectors. From
these solutions we reconstruct the average volume function, the scalar
fields, the directional anisotropic functions, the normalized Hubble
parameter, the deceleration parameter, and the effective
equation-of-state parameter. We show that, although the different
branches display distinct transient cosmological evolutions, all
physically admissible solutions converge toward a common asymptotic de
Sitter state characterized by
$q\rightarrow-1$, $H\rightarrow H_0$, and
$w_{\rm eff}\rightarrow-1$.

The overall analytical strategy adopted in this work is summarized in
Fig.~\ref{fig:flowchart}. Starting from the generalized action, a field
redefinition and a canonical transformation lead to the Hamiltonian
formulation of the model. The elimination of the mixed kinetic
contribution imposes a constraint on $m_{12}^{\pm}$, giving rise to
three Hamiltonian structures and six exact cosmological branches. These
solutions determine the average volume function, the scalar fields, and
the directional anisotropic functions, allowing the reconstruction of
the corresponding cosmological observables.

The paper is organized as follows. In Section II we formulate the
generalized Saez--Ballester--K-essence-like theory within the framework
of chiral cosmology and derive the corresponding field equations. We
also obtain the constraint on the mixed kinetic coupling, which
determines the admissible cosmological branches. Section III introduces
the Hamiltonian formulation and identifies the three Hamiltonian
structures from which the exact solutions are constructed. Sections
IV--VI are devoted to the exact analytical solutions describing the
quintessence, phantom, and quintom scalar-field sectors, respectively,
where both branches associated with the $m_{12}^{\pm}$ constraint are
analyzed. In Section VII we reconstruct the cosmological observables
associated with the exact solutions. Besides the average volume
function and the scalar fields, we discuss the directional scale
functions that encode the anisotropic nature of the Bianchi type I
geometry and reconstruct the average Hubble parameter, the deceleration
parameter, and the effective equation-of-state parameter. Section VIII
presents a comparative analysis of the six exact cosmological branches,
highlighting their common asymptotic de Sitter attractor and their
distinct transient cosmological evolutions. Section IX discusses the
physical interpretation of the exact solutions, including the origin of
the universal late-time de Sitter behavior, the role of the chiral
multifield structure and inverse power-law potentials, and the meaning
of the reconstructed observables in an anisotropic Bianchi type I
spacetime. Finally, Section X summarizes the main results and outlines
possible directions for future research.

\section{Chiral cosmology in the Saez--Ballester--K-essence-like theory}

We consider a generalized Saez--Ballester--K-essence-like theory
containing two interacting scalar fields with a noncanonical chiral
kinetic sector. The scalar fields are assumed to interact through a
power-law self-interaction potential, while the kinetic coupling is
described by a field-dependent chiral matrix
$m^{ab}(\psi_i)$. The corresponding action is
\begin{equation}
{\cal L}=\sqrt{-g}\left(
R-\frac12 g^{\mu\nu}
m^{ab}(\psi_i)\,
\nabla_\mu\psi_a
\nabla_\nu\psi_b
+
V(\psi_1,\psi_2)
\right),
\label{lagra}
\end{equation}
where $R$ is the Ricci scalar and
\begin{equation}
V(\psi_1,\psi_2)
=
V_1\psi_1^{\pm\lambda_1}
+
V_2\psi_2^{\pm\lambda_2},
\end{equation}
is the scalar-field self-interaction potential. Throughout this work
the parameters $\lambda_i>0$ determine the power-law exponents, and
both positive- and negative-power potentials are considered. The
kinetic interaction is encoded in the symmetric chiral matrix
\begin{equation}
m^{ab}=
\left(
\begin{array}{cc}
\dfrac{m^{11}}{\psi_1^2}
&
\dfrac{m^{12}}{\psi_1\psi_2}
\\[0.4cm]
\dfrac{m^{21}}{\psi_1\psi_2}
&
\dfrac{m^{22}}{\psi_2^2}
\end{array}
\right),
\qquad
m^{12}=m^{21}.
\end{equation}
The diagonal components determine the kinetic character of the scalar
sector. In particular, $m^{11}=m^{22}=+1$ corresponds to two quintessence fields, $m^{11}=m^{22}=-1$ describes two phantom fields, whereas $m^{11}=+1$,  $m^{22}=-1$, or vice versa, defines a mixed (quintom) kinetic sector.

To simplify the field equations we introduce the field redefinition $\psi_i=e^{\phi_i},$ with $i=1,2,$ under which the action becomes
\begin{equation}
{\cal L}
=
\sqrt{-g}
\left(
R
-
\frac12
g^{\mu\nu}
m^{ab}
\nabla_\mu\phi_a
\nabla_\nu\phi_b
+
V(\phi_1,\phi_2)
\right),
\label{lagra_phi}
\end{equation}
where
\begin{equation}
V(\phi_1,\phi_2)
=
V_1e^{\pm\lambda_1\phi_1}
+
V_2e^{\pm\lambda_2\phi_2},
\end{equation}
is now an exponential scalar-field potential, while the matrix $m^{ab}$ becomes constant.

Varying the action (\ref{lagra_phi}) with respect to the metric and
the scalar fields yields the Einstein--Klein--Gordon field equations,
\begin{equation}
G_{\alpha\beta}
=
-\frac12
m^{ab}
\left(
\nabla_\alpha\phi_a
\nabla_\beta\phi_b
-
\frac12
g_{\alpha\beta}
g^{\mu\nu}
\nabla_\mu\phi_a
\nabla_\nu\phi_b
\right)
+
\frac12
g_{\alpha\beta}
V(\phi_1,\phi_2),
\label{mono}
\end{equation}
and
\begin{equation}
m^{ab}\Box\phi_b
-
\frac{\partial V}{\partial\phi_a}
=
m^{ab}
g^{\mu\nu}
{\phi_b}_{,\mu\nu}
-
m^{ab}
g^{\alpha\beta}
\Gamma^\nu_{\alpha\beta}
\nabla_\nu\phi_b
-
\frac{\partial V}{\partial\phi_a}
=
0,
\qquad
a,b=1,2,
\label{ekg-phi}
\end{equation}
which constitute the starting point for the Hamiltonian formulation
developed below.
We consider the line element for the anisotropic cosmological
Bianchi type I model in the Misner's parametrization
\begin{eqnarray}
\rm ds^2 &=& \rm -N^2 dt^2 +a_1^2 dx^2 + a_2^2 dy^2 + a_3^2 dz^2, \nonumber \\
 &=&\rm -N^2 dt^2 + e^{2\Omega}\left[e^{2\beta_++2\sqrt{3}\beta_-}dx^2 +e^{2\beta_+-2\sqrt{3}\beta_-} dy^2 + e^{-4\beta_+} dz^2\right],
 \label{biachi_I_misner}
 \end{eqnarray}
 where $\rm a_i$ ($\rm i=1,2,3$) are the scale factor on directions $\rm (x,y,z)$, respectively, and N is the lapse function.
 For convenience, and in order to carry out the analytical calculations, we consider the following representation for the line
 element (\ref{biachi_I_misner})
\begin{equation}
\rm ds^2 = -N^2 dt^2 + \eta^2\left[m_1^2 dx^2 +m_2^2 dy^2 + m_3^2
dz^2\right], \label{bianchi}
 \end{equation}
 where the relations between both representations are given by
 \begin{eqnarray}
 \rm \eta&=& \rm e^{\Omega}, \nonumber\\
 \rm m_1 &=& \rm e^{\beta_+ + \sqrt{3} \beta_-}, \qquad \frac{\dot m_1}{m_1}=\dot \beta_+ + \sqrt{3} \dot \beta_-, \nonumber\\
 \rm m_2&=& \rm e^{\beta_+ - \sqrt{3} \beta_-}, \qquad \frac{\dot m_2}{m_2}=\dot \beta_+ - \sqrt{3} \dot \beta_-,\nonumber\\
 \rm m_3&=& \rm e^{-2\beta_+ }, \qquad \qquad \frac{\dot m_3}{m_3}=-2\dot \beta_+, \label{rel}
\end{eqnarray}
 where $\eta$ is a function that has information regarding the isotropic scenario and the $\rm m_i$ are dimensionless functions
 that has  information about the anisotropic behavior of the universe, such that
\begin{equation}
\rm  \Pi_{i=1}^{3}m_i=1, \qquad
 \Pi_{i=1}^{3}a_i= \eta^3, \qquad
\rm  \sum_{i=1}^3 \frac{\dot m_i}{m_i}= 0, \label{mi}
\end{equation}
act as constraint equations for the model.

Throughout this work we focus on inverse power-law scalar potentials,
\[
V(\phi_i)=V_1e^{-\lambda_1\phi_1}
+V_2e^{-\lambda_2\phi_2},
\]
whereas the corresponding positive-power case will be investigated
elsewhere.

Substituting the metric (\ref{bianchi}) into the action
(\ref{lagra_phi}), together with the relations (\ref{rel}), the minisuperspace Lagrangian takes the form
\begin{eqnarray}\label{lagra-mixto}
{\cal L}
&=&
\eta^3
\left(
\frac{6}{N}\left(\frac{\dot\eta}{\eta}\right)^2
-\frac{1}{N}
\left[
\left(
\frac{\dot m_1}{m_1}
\right)^2
+
\left(
\frac{\dot m_2}{m_2}
\right)^2
+
\left(
\frac{\dot m_3}{m_3}
\right)^2
\right]
\right.
\nonumber\\
&&
\left.
-
m^{11}\frac{\dot\phi_1^2}{2N}
-
m^{22}\frac{\dot\phi_2^2}{2N}
-
m^{12}\frac{\dot\phi_1\dot\phi_2}{N}
+
N
\left[
V_1e^{-\lambda_1\phi_1}
+
V_2e^{-\lambda_2\phi_2}
\right]
\right),
\end{eqnarray}
where the canonical momenta,
\[
\Pi_q=\frac{\partial{\cal L}}{\partial\dot q},
\]
are given by
\begin{eqnarray}
\Pi_\eta
&=&
12\frac{\eta}{N}\dot\eta,
\qquad\qquad
\dot\eta
=
\frac{N}{12\eta}\Pi_\eta,
\nonumber\\
\Pi_{\phi_1}
&=&
-\frac{\eta^3}{N}
\left(
m^{11}\dot\phi_1
+
m^{12}\dot\phi_2
\right),
\qquad
\dot\phi_1
=
\frac{N}{\eta^3\Delta}
\left(
-m^{22}\Pi_{\phi_1}
+
m^{12}\Pi_{\phi_2}
\right),
\label{momenta-m}
\\
\Pi_{\phi_2}
&=&
-\frac{\eta^3}{N}
\left(
m^{22}\dot\phi_2
+
m^{12}\dot\phi_1
\right),
\qquad
\dot\phi_2
=
\frac{N}{\eta^3\Delta}
\left(
m^{12}\Pi_{\phi_1}
-
m^{11}\Pi_{\phi_2}
\right),
\nonumber\\
\Pi_1
&=&
-\frac{2\eta^3}{N}
\left(
\frac{\dot m_1}{m_1^2}
\right),
\qquad\qquad
\dot m_1
=
-\frac{Nm_1^2\Pi_1}{2\eta^3},
\nonumber\\
\Pi_2
&=&
-\frac{2\eta^3}{N}
\left(
\frac{\dot m_2}{m_2^2}
\right),
\qquad\qquad
\dot m_2
=
-\frac{Nm_2^2\Pi_2}{2\eta^3},
\nonumber\\
\Pi_3
&=&
-\frac{2\eta^3}{N}
\left(
\frac{\dot m_3}{m_3^2}
\right),
\qquad\qquad
\dot m_3
=
-\frac{Nm_3^2\Pi_3}{2\eta^3}.
\nonumber
\end{eqnarray} 
here $\Delta=m^{11}m^{22}-(m^{12})^2$.
Expressing the Lagrangian in canonical form,
\begin{equation}
    {\cal L}_{\rm can}
=
\Pi_q\dot q
-
N{\cal H},
\end{equation}
and varying it with respect to the lapse function,
\begin{equation}
    \frac{\delta{\cal L}_{\rm can}}{\delta N}=0,
\end{equation}
one obtains the Hamiltonian constraint, ${\cal H}=0,$ with Hamiltonian density
\begin{eqnarray}
{\cal H}
&=&
\frac{1}{24\eta^3}
\left[
\eta^2\Pi_\eta^2
-
\frac{12}{\Delta}
\left(
m^{22}\Pi_{\phi_1}^2
-
2m^{12}\Pi_{\phi_1}\Pi_{\phi_2}
+
m^{11}\Pi_{\phi_2}^2
\right)
\right.
\nonumber\\
&&
\left.
-
6
\left(
m_1^2\Pi_1^2
+
m_2^2\Pi_2^2
+
m_3^2\Pi_3^2
\right)
-
24V_1\eta^6e^{-\lambda_1\phi_1}
-
24V_2\eta^6e^{-\lambda_2\phi_2}
\right].
\label{hamiltonian}
\end{eqnarray}
To simplify the Hamiltonian structure we introduce the canonical
transformation
\begin{equation}
    \Pi_\eta=\frac{\partial S}{\partial\eta},
\qquad
\Pi_i=\frac{\partial S}{\partial m_i},
\end{equation}
together with the variables $\eta=e^u$, $m_i=e^{u_i}$, where
\begin{equation}
    P_i=\frac{\partial S}{\partial u_i},
\qquad
\Pi_u=\frac{\partial S}{\partial u}.
\end{equation}
Under these transformations, the Hamiltonian density becomes
{\small
\begin{equation}
{\cal H}
=
\frac{e^{-3u}}{24}
\left[
\Pi_u^2
-
6P_1^2
-
6P_2^2
-
6P_3^2
-
\frac{12m^{22}}{\Delta}\Pi_{\phi_1}^2
-
\frac{12m^{11}}{\Delta}\Pi_{\phi_2}^2
+
\frac{24m^{12}}{\Delta}
\Pi_{\phi_1}\Pi_{\phi_2}
-
24V_1e^{6u-\lambda_1\phi_1}
-
24V_2e^{6u-\lambda_2\phi_2}
\right].
\label{hami-exponential}
\end{equation}
}

To simplify the Hamiltonian structure we introduce the following
canonical transformation between the variables
$(u,\phi_1,\phi_2,u_i)$ and
$(\xi_1,\xi_2,\xi_3,u_i)$,
\begin{equation}\label{trans_2}
\begin{split}
\xi_1&=-6u+\lambda_1\phi_1,\\
\xi_2&=-6u+\lambda_2\phi_2,\\
\xi_3&=-4u+\frac{\lambda_1}{6}\phi_1
+\frac{\lambda_2}{6}\phi_2,\\
u_i&=u_i,
\end{split}
\qquad\Longleftrightarrow\qquad
\begin{split}
u&=\frac{\xi_1+\xi_2-6\xi_3}{12},\\
\phi_1&=\frac{3\xi_1+\xi_2-6\xi_3}{2\lambda_1},\\
\phi_2&=\frac{\xi_1+3\xi_2-6\xi_3}{2\lambda_2}.
\end{split}
\end{equation}
Choosing the gauge
\[
N=24e^{3u},
\]
the corresponding canonical momenta become
\begin{align}
\Pi_u&=-6\pi_1-6\pi_2-4\pi_3,\nonumber\\
\Pi_{\phi_1}&=\lambda_1\pi_1+\frac{\lambda_1}{6}\pi_3,
\nonumber\\
\Pi_{\phi_2}&=\lambda_2\pi_2+\frac{\lambda_2}{6}\pi_3.
\label{new-moment}
\end{align}
Under these transformations, the Hamiltonian density takes the form
\begin{align}
{\cal H}
&=
12\left(3-\frac{\lambda_1^2m^{22}}{\triangle}\right)\pi_1^2
+
12\left(3-\frac{\lambda_2^2m^{11}}{\triangle}\right)\pi_2^2
\nonumber\\
&
+
\left(
16+
\frac{-\lambda_1^2m^{22}
+2\lambda_1\lambda_2m^{12}
-\lambda_2^2m^{11}}
{3\triangle}
\right)\pi_3^2
\nonumber\\
&
+
12
\left[
\left(
4+
\frac{\lambda_1\lambda_2m^{12}
-\lambda_1^2m^{22}}
{3\triangle}
\right)\pi_1
+
\left(
4+
\frac{\lambda_1\lambda_2m^{12}
-\lambda_2^2m^{11}}
{3\triangle}
\right)\pi_2
\right]
\pi_3
\nonumber\\
&
-6P_1^2
-6P_2^2
-6P_3^2
+
24
\left(
3+\frac{\lambda_1\lambda_2m^{12}}{\triangle}
\right)
\pi_1\pi_2
-
24
\left(
V_1e^{-\xi_1}
+
V_2e^{-\xi_2}
\right),
\label{hamiltonian}
\end{align}
where $\triangle$ is defined in Eq.~(\ref{momenta-m}). The exact analytical solutions are obtained from Hamilton's equations,
\begin{eqnarray}
\dot\xi_1
&=&
24
\left(
3-\frac{\lambda_1^2m^{22}}{\triangle}
\right)\pi_1
+
24
\left(
3+\frac{\lambda_1\lambda_2m^{12}}{\triangle}
\right)\pi_2
\nonumber\\
&&
+
12
\left(
4+
\frac{\lambda_1\lambda_2m^{12}
-\lambda_1^2m^{22}}
{3\triangle}
\right)\pi_3,
\nonumber\\
\dot\xi_2
&=&
24
\left(
3-\frac{\lambda_2^2m^{11}}{\triangle}
\right)\pi_2
+
24
\left(
3+\frac{\lambda_1\lambda_2m^{12}}{\triangle}
\right)\pi_1
\nonumber\\
&&
+
12
\left(
4+
\frac{\lambda_1\lambda_2m^{12}
-\lambda_2^2m^{11}}
{3\triangle}
\right)\pi_3,
\nonumber\\
\dot\xi_3
&=&
12
\left[
\left(
4+
\frac{\lambda_1\lambda_2m^{12}
-\lambda_1^2m^{22}}
{3\triangle}
\right)\pi_1
+
\left(
4+
\frac{\lambda_1\lambda_2m^{12}
-\lambda_2^2m^{11}}
{3\triangle}
\right)\pi_2
\right]
\nonumber\\
&&
+
2
\left(
16+
\frac{-\lambda_1^2m^{22}
+2\lambda_1\lambda_2m^{12}
-\lambda_2^2m^{11}}
{3\triangle}
\right)\pi_3,
\nonumber\\
\dot\pi_1
&=&
-24V_1e^{-\xi_1},
\label{ecs-mov-2}
\\
\dot\pi_2
&=&
-24V_2e^{-\xi_2},
\nonumber\\
\dot\pi_3
&=&
0,
\qquad
\dot P_i=0,
\qquad
\dot u_i=-12P_i.
\nonumber
\end{eqnarray}
The last three equations immediately imply that
\begin{equation}
    \pi_3=p_3,
\qquad
P_i=n_i,
\end{equation}
where $p_3$ and $n_i$ are integration constants. Consequently,
\begin{equation}
    u_i=u_{i0}-12n_i\Delta t.
\end{equation}
Taking the time derivative of the first equation in
Eq.~(\ref{ecs-mov-2}) yields
\begin{equation}
\ddot\xi_1=
-576V_1
\left(
3-\frac{\lambda_1^2m^{22}}{\triangle}
\right)
e^{-\xi_1}
-
576V_2
\left(
3+\frac{\lambda_1\lambda_2m^{12}}{\triangle}
\right)
e^{-\xi_2}.
\label{first}
\end{equation}
The canonical transformation (\ref{trans_2}) has been introduced so
that the Hamiltonian can be diagonalized after imposing an appropriate
constraint on the mixed momentum term in
Eq.~(\ref{hamiltonian}). Requiring the coefficient of
$\pi_1\pi_2$ to vanish leads to the condition
\begin{equation}
\label{m12}
m^{12}
=
\frac{\lambda_1\lambda_2}{6}
\left(
1
\pm
\sqrt{
1+
36
\frac{m^{11}m^{22}}
{\lambda_1^2\lambda_2^2}
}
\right).
\end{equation}
The reality of the square root requires
\begin{equation}
    1+
36
\frac{m^{11}m^{22}}
{\lambda_1^2\lambda_2^2}
\ge0,
\end{equation}
while the above constraint is valid for
$\lambda_1\neq\lambda_2$ and
$\lambda_i\ge2$ in the quintessence and phantom sectors.

The condition (\ref{m12}) naturally separates the theory into three
physically distinct kinetic sectors, namely quintessence, phantom, and
quintom. Each sector admits two exact cosmological branches
corresponding to the two possible signs in Eq.~(\ref{m12}). These
branches are analyzed separately in the following sections.

\begin{figure}[h!]
\centering
\begin{tikzpicture}[
node distance=1.05cm,
every node/.style={
    rectangle,
    rounded corners,
    draw=black,
    align=center,
    minimum width=5.8cm,
    minimum height=0.75cm,
    font=\small
},
arrow/.style={->, thick}
]

\node (action) {Generalized Saez--Ballester--K-essence-like action};
\node (transform) [below of=action] {Field redefinition and canonical transformation};
\node (hamiltonian) [below of=transform] {Hamiltonian formulation};
\node (constraint) [below of=hamiltonian] {Constraint on the mixed kinetic term $m_{12}^{\pm}$};
\node (classes) [below of=constraint] {Three Hamiltonian structures};
\node (solutions) [below of=classes] {Six exact cosmological solutions};
\node (functions) [below of=solutions] {$\eta^3(t)$, $\psi_1(t)$, $\psi_2(t)$};
\node (observables) [below of=functions] {$H(t)$, $q(t)$, $w_{\rm eff}(t)$};
\node (physics) [below of=observables] {Transient dynamics and common de Sitter attractor};

\draw[arrow] (action) -- (transform);
\draw[arrow] (transform) -- (hamiltonian);
\draw[arrow] (hamiltonian) -- (constraint);
\draw[arrow] (constraint) -- (classes);
\draw[arrow] (classes) -- (solutions);
\draw[arrow] (solutions) -- (functions);
\draw[arrow] (functions) -- (observables);
\draw[arrow] (observables) -- (physics);

\end{tikzpicture}
\caption{
Flowchart illustrating the analytical framework developed in this work,
from the generalized SBK like action to the
identification of the universal late-time de Sitter attractor. A field
redefinition and a canonical transformation lead to the Hamiltonian
formulation of the model. The elimination of the mixed kinetic
contribution imposes the constraint $m_{12}^{\pm}$, giving rise to
three Hamiltonian structures and six exact cosmological branches. The
corresponding analytical solutions determine the average volume
function, the scalar fields, and the directional anisotropic
functions, enabling the reconstruction of the average cosmological
observables $H(t)$, $q(t)$, and $w_{\rm eff}(t)$. Although the six
branches exhibit distinct transient cosmological evolutions, they all
converge toward a common asymptotic de Sitter attractor.
}
\label{fig:flowchart}
\end{figure}

\newpage

\section{Hamiltonian density structures and resulting dynamics}
In our analysis presented below, the six cases can be described by three different Hamiltonian structure respect to the $\pi_i$-momentum sign.

\subsubsection{Structure A}
The first case of Hamiltonian density has the form
\begin{equation}
\rm {\cal H}= -q^a_{_\ell}\pi_1^2 -q^b_{_\ell}\pi_2^2
+c_{_\ell}\pi_3^2 +\left(d^a_{_\ell}\pi_1+ d^b_{_\ell}\pi_2 \right)
\pi_3-6\left(P_1^2+P_2^2+P_3^2\right)-24V_1e^{-\xi_1}-24V_2
e^{-\xi_2}, \label{hami-A}
\end{equation}
thus, Hamilton equations for the new simplified coordinate $\rm \xi_i$, (\ref{ecs-mov-2}), are
\begin{eqnarray}\label{new_ecs_mov_1}
\rm \dot \xi_1&=& \rm -2q^a_{_\ell}\pi_1
+d^a_{_\ell}\pi_3, \nonumber\\
\rm \dot \xi_2&=& \rm -2{q^b_{_\ell}}\pi_2
+d^b_{_\ell}\pi_3, \\
\rm \dot \xi_3&=& \rm 2 c_{_\ell}\pi_3+
d^a_{_\ell}\pi_1+d^b_{_\ell}\pi_2  , \nonumber 
\end{eqnarray}
the equations for $\rm \dot \pi_i$ remain the same as in
(\ref{ecs-mov-2}). Taking the derivative of the first equation of
(\ref{new_ecs_mov_1}) yields
\begin{equation}
\rm \ddot \xi_1= 48 V_1 q^a_{_\ell} e^{-\xi_1},
\end{equation}
which has a solution of the form
\begin{equation}
\rm e^{-\xi_1}=\frac{ r_1^2}{24 V_1 q^a_{_\ell}} \, Sech^2\left(r_1
t-q_1\right). \label{solucion-xi1}
\end{equation}
From Eq.~(\ref{new_ecs_mov_2}), it is evident that $\dot{\xi}_2$ has the same functional structure as $\dot{\xi}_1$, implying that its solution has the same functional form as Eq.~(\ref{solucion-xi1}), so we have
\begin{equation}
\rm e^{-\xi_2}=\frac{ r_2^2}{24 V_2 q^b_{_\ell}} \, Sech^2\left(r_2
t-q_2\right), \label{solucion-xi2}
\end{equation}
where $\rm r_i$ and $\rm q_i$ (with $\rm i=1,2$) are integration
constants, both at (\ref{solucion-xi1}) and (\ref{solucion-xi2}).
Reinserting these solutions into Hamilton equations for the momenta,
we obtain
\begin{eqnarray}
\rm \pi_1 &=& \rm a_1 - \frac{r_1}{q^a_{_\ell}} \, Tanh\left(r_1
t-q_1
\right), \label{solution-p1}\\
\rm \pi_2 &=& \rm a_2 - \frac{r_2}{q^b_{_\ell}} \, Tanh\left(r_2
t-q_2 \right), \label{solucion-p2}
\end{eqnarray}
With (\ref{solution-p1}) and (\ref{solucion-p2}), it can be easily
checked that the Hamiltonian is identically null when
\begin{equation}
\rm a_1=\frac{d^a_{_\ell}}{2q^a_{_\ell}}p_3,\qquad
a_2=\frac{d^b_{_\ell}}{2q^b_{_\ell}}p_3,\qquad
p_3^2=4\frac{6q^a_{_\ell} q^b_{_\ell}n^2 +q^b_{_\ell}r_1^2+
q^a_{_\ell}r_2^2}{4c_{_\ell}
q^a_{_\ell}q^b_{_\ell}+q^b_{_\ell}d^{a2}_{_\ell}+q^a_{_\ell}d^{b2}_{_\ell}},
\qquad n^2=n_1^2+n_2^2+n_3^2.
\end{equation}
where $\rm n^2$ represents the contribution from the anisotropic
functions. We are now in position to write the solutions for the $\rm
\xi_i$ coordinates, which read
\begin{eqnarray}
\rm \xi_1 &=& \rm \beta_1 +Ln\left[ Cosh^2\left(r_1 t -q_1 \right)\right],  \label{xi1}\\
\xi_2 &=& \rm \beta_2+Ln\left[ Cosh^2\left(r_2 t -q_2 \right)\right], \label{xi2}\\
\rm \xi_3&=&\rm \beta_3 +
\frac{p_3}{2q^a_{_\ell}q^b_{_\ell}}\left[4c_{_\ell}q^a_{_\ell}q^b_{_\ell}+q^b_{_\ell}d^{a2}_{_\ell}+q^a_{_\ell}d^{b2}_{_\ell}\right]\Delta
t-\frac{d^a_{_\ell}}{q^a_{_\ell}}\, Ln\,\left[Cosh\left(r_1t-q_1
\right) \right]-\frac{d^b_{_\ell}}{q^b_{_\ell}}\,
Ln\,\left[Cosh\left(r_2 t-q_2 \right) \right],
\end{eqnarray}

here the $\rm \beta_i$, with $\rm i=1,2,3$, arise as integration constants . Applying the inverse canonical
transformation we obtain the solutions in the original variables
$\rm (\eta, \phi_1, \phi_2)$ as
\begin{eqnarray}
\eta&=&e^u=\eta_0\,
Exp\left[-(q^b_{_\ell}d^{a2}_{_\ell}+q^a_{_\ell}d^{b2}_{_\ell}+4c_{_\ell}q^a_{_\ell}q^b_{_\ell})\frac{p_3}{4q^a_{_\ell}q^b_{_\ell}}\Delta
t\right] \, \left[Cosh(r_1t-q_1)\right]^{a^a_{_\ell}}
\left[Cosh(r_2t-q_2)\right]^{a^b_{_\ell}},\label{etan1}\\
\phi_1&=& \phi_{10} -\frac{3p_3}{2\lambda_1 \,
q^a_{_\ell}q^b_{_\ell}}\left(4c_{_\ell}q^a_{_\ell}q^b_{_\ell}+q^b_{_\ell}d^{a2}_{_\ell}
+q^a_{_\ell}d^{b2}_{_\ell} \right)\Delta t
+Ln\left[Cosh^{\theta_1}\left(r_1t-q_1 \right) \, Cosh^{\theta_2}\left(r_2t-q_2 \right)\right]\\
\phi_2&=& \phi_{20} - \frac{3p_3}{2\lambda_2 \,
q^a_{_\ell}q^b_{_\ell}}\left[4c_{_\ell}q^a_{_\ell}q^b_{_\ell}+d^{a2}_{_\ell}q^b_{_\ell}+
d^{b2}_{_\ell}q^a_{_\ell} \right]\Delta t
+Ln\left[Cosh^{\theta_3}\left(r_1t-q_1 \right) \,
Cosh^{\theta_4}\left(r_2t-q_2 \right)\right] \label{sols_quint}
\end{eqnarray}

where $\eta_0, \phi_{10}$ and $\phi_{20}$ are given in terms of the
$\beta_i$ constants as
\begin{equation}
\eta_0=e^{\frac{\beta_1+\beta_2-6\beta_3}{12}}, \quad
\phi_{10}=\frac{3\beta_1+\beta_2-6\beta_3}{2\lambda_1},\quad
\phi_{20}=\frac{\beta_1+3\beta_2-6\beta_3}{2\lambda_2}.\label{phi2}
\end{equation}
 and the constants $a^a_{_\ell}=\frac{q^a_{_\ell}+3d^a_{_\ell}}{6q^a_{_\ell}}$,
$a^b_{_\ell}=\frac{q^b_{_\ell}+3d^b_{_\ell}}{6q^b_{_\ell}}$,
$\theta_1=\frac{3q^a_{_\ell}+3d^a_{_\ell}}{\lambda_1q^a_{_\ell}}$,
$\theta_2=\frac{q^b_{_\ell}+3d^b_{_\ell}}{\lambda_1 q^b_{_\ell}}$,
$\theta_3=\frac{q^a_{_\ell}+3d^a_{_\ell}}{\lambda_2\,q^a_{_\ell}}$,
$\theta_4=\frac{3
q^b_{_\ell}+3d^b_{_\ell}}{\lambda_2\,q^b_{_\ell}}$.

 Finally, the volume function and the scalar function
$\psi_i$ become
\begin{eqnarray}
\eta^3&=&\eta_0^3\,
Exp\left[-(q^b_{_\ell}d^{a2}_{_\ell}+q^a_{_\ell}d^{b2}_{_\ell}+4c_{_\ell}q^a_{_\ell}q^b_{_\ell})\frac{3p_3}{4q^a_{_\ell}q^b_{_\ell}}\Delta
t\right] \, \left[Cosh(r_1t-q_1)\right]^{3a^a_{_\ell}}
\left[Cosh(r_2t-q_2)\right]^{3a^b_{_\ell}},\label{eta_A}\\
\psi_1&=& \psi_{10} \,Exp\left[-\frac{3p_3}{2\lambda_1 \,
q^a_{_\ell}q^b_{_\ell}}\left(4c_{_\ell}q^a_{_\ell}q^b_{_\ell}+q^b_{_\ell}d^{a2}_{_\ell}
+q^a_{_\ell}d^{b2}_{_\ell} \right)\Delta t\right]
\,\left[Cosh^{\theta_1}\left(r_1t-q_1 \right) \, Cosh^{\theta_2}\left(r_2t-q_2 \right)\right]\label{psi1_A}\\
\psi_2&=& \psi_{20}\,Exp\left[- \frac{3p_3}{2\lambda_2 \,
q^a_{_\ell}q^b_{_\ell}}\left[4c_{_\ell}q^a_{_\ell}q^b_{_\ell}+d^{a2}_{_\ell}q^b_{_\ell}+
d^{b2}_{_\ell}q^a_{_\ell} \right]\Delta t \right]
\left[Cosh^{\theta_3}\left(r_1t-q_1 \right) \,
Cosh^{\theta_4}\left(r_2t-q_2 \right)\right] \label{psi2_A}
\end{eqnarray}
where $q^a_{_\ell}, \, q^b_{_\ell}, \, d^{a}_{_\ell}, \, d^{b}_{_\ell}, \, c_{_\ell}, \, a^a_{_\ell}, \,$ and $ a^b_{_\ell} $ are parameters determined for each case.

\subsubsection{Structure B}
For this case, we have 
\begin{equation}
\rm {\cal H}= q^a_{_\ell} \pi_1^2 +q^b_{_\ell}\pi_2^2
+c_{_\ell}\pi_3^2 +\left(d^a_{_\ell}\pi_1+d^b_{_\ell}\pi_2 \right)
\pi_3-6\left(P_1^2+P_2^2+P_3^2\right)-24V_1e^{-\xi_1}-24V_2
e^{-\xi_2}, \label{hami-B}
\end{equation}
thus, Hamilton equations for the new simplified coordinate $\rm \xi_i$ are
\begin{eqnarray}\label{new_ecs_mov_2}
\rm \dot \xi_1&=& \rm -2q^a_{_\ell}\pi_1
+d^a_{_\ell}\pi_3, \nonumber\\
\rm \dot \xi_2&=& \rm -2{q^b_{_\ell}}\pi_2
+d^b_{_\ell}\pi_3, \\
\rm \dot \xi_3&=& \rm 2 c_{_\ell}\pi_3+
d^a_{_\ell}\pi_1+d^b_{_\ell}\pi_2  , \nonumber
\end{eqnarray}
the equations for $\rm \dot \pi_i$ remain the same as in
(\ref{ecs-mov-2}). Taking the derivative of the first equation of
(\ref{new_ecs_mov_2}) yields
\begin{equation}
\rm \ddot \xi_1= 48 V_1 q^a_{_\ell} e^{-\xi_1},
\end{equation}
which has a solution of the form
\begin{equation}
\rm e^{-\xi_1}=\frac{ r_1^2}{24 V_1 q^a_{_\ell}} \, Sech^2\left(r_1
t-q_1\right). \label{solucion-xi1}
\end{equation}
From Eq.~(\ref{new_ecs_mov_2}), it is evident that $\dot{\xi}_2$ has the same functional structure as $\dot{\xi}_1$. Its solution therefore follows directly from Eq.~(\ref{solucion-xi1}) and can be written as
\begin{equation}
\rm e^{-\xi_2}=\frac{ r_2^2}{24 V_2 q^b_{_\ell}} , \mathrm{sech}^2\left(r_2
t-q_2\right),
\label{solucion-xi2}
\end{equation}

where $\rm r_i$ and $\rm q_i$ (with $\rm i=1,2$) are integration
constants, both at (\ref{solucion-xi1}) and (\ref{solucion-xi2}).
Reinserting these solutions into Hamilton equations for the momenta,
we obtain
\begin{eqnarray}
\rm \pi_1 &=& \rm a_1 - \frac{r_1}{q^a_{_\ell}} \, Tanh\left(r_1
t-q_1
\right), \label{solution-p1}\\
\rm \pi_2 &=& \rm a_2 - \frac{r_2}{q^b_{_\ell}} \, Tanh\left(r_2
t-q_2 \right), \label{solucion-p2}
\end{eqnarray}
With (\ref{solution-p1}) and (\ref{solucion-p2}), one can  easily verify that the Hamiltonian is identically null when
\begin{equation}
\rm a_1=\frac{d^a_{_\ell}}{2q^a_{_\ell}}p_3,\qquad
a_2=\frac{d^b_{_\ell}}{2q^b_{_\ell}}p_3,\qquad
p_3^2=4\frac{6q^a_{_\ell} q^b_{_\ell}n^2 +q^b_{_\ell}r_1^2+
q^a_{_\ell}r_2^2}{4c_{_\ell}
q^a_{_\ell}q^b_{_\ell}+q^b_{_\ell}d^{a2}_{_\ell}+q^a_{_\ell}d^{b2}_{_\ell}},
\qquad n^2=n_1^2+n_2^2+n_3^2.
\end{equation}
where $\rm n^2$ corresponds to the contribution on the anisotropic
functions.  We are now in a position to write the solutions for the $\rm
\xi_i$ coordinates, which read
\begin{eqnarray}
\rm \xi_1 &=& \rm \beta_1 +Ln\left[ Cosh^2\left(r_1 t -q_1 \right)\right],  \label{xi1}\\
\xi_2 &=& \rm \beta_2+Ln\left[ Cosh^2\left(r_2 t -q_2 \right)\right], \label{xi2}\\
\rm \xi_3&=&\rm \beta_3 +
\frac{p_3}{2q^a_{_\ell}q^b_{_\ell}}\left[4c_{_\ell}q^a_{_\ell}q^b_{_\ell}+q^b_{_\ell}d^{a2}_{_\ell}+q^a_{_\ell}d^{b2}_{_\ell}\right]\Delta
t-\frac{d^a_{_\ell}}{q^a_{_\ell}}\, Ln\,\left[Cosh\left(r_1t-q_1
\right) \right]-\frac{d^b_{_\ell}}{q^b_{_\ell}}\,
Ln\,\left[Cosh\left(r_2 t-q_2 \right) \right],
\end{eqnarray}

here,  the constants $\rm \beta_i\,(i=1,2,3)$ arise as constants
coming from integration. Applying the inverse canonical
transformation we obtain the solutions in the original variables
$\rm (\eta, \phi_1, \phi_2)$ as
\begin{eqnarray}
\eta&=&e^u=\eta_0\,
Exp\left[-(q^b_{_\ell}d^{a2}_{_\ell}+q^a_{_\ell}d^{b2}_{_\ell}+4c_{_\ell}q^a_{_\ell}q^b_{_\ell})\frac{p_3}{4q^a_{_\ell}q^b_{_\ell}}\Delta
t\right] \, \left[Cosh(r_1t-q_1)\right]^{a^a_{_\ell}}
\left[Cosh(r_2t-q_2)\right]^{a^b_{_\ell}},\label{etan1}\\
\phi_1&=& \phi_{10} -\frac{3p_3}{2\lambda_1 \,
q^a_{_\ell}q^b_{_\ell}}\left(4c_{_\ell}q^a_{_\ell}q^b_{_\ell}+q^b_{_\ell}d^{a2}_{_\ell}
+q^a_{_\ell}d^{b2}_{_\ell} \right)\Delta t
+Ln\left[Cosh^{\theta_1}\left(r_1t-q_1 \right) \, Cosh^{\theta_2}\left(r_2t-q_2 \right)\right]\\
\phi_2&=& \phi_{20} - \frac{3p_3}{2\lambda_2 \,
q^a_{_\ell}q^b_{_\ell}}\left[4c_{_\ell}q^a_{_\ell}q^b_{_\ell}+d^{a2}_{_\ell}q^b_{_\ell}+
d^{b2}_{_\ell}q^a_{_\ell} \right]\Delta t
+Ln\left[Cosh^{\theta_3}\left(r_1t-q_1 \right) \,
Cosh^{\theta_4}\left(r_2t-q_2 \right)\right] \label{sols_quint}
\end{eqnarray}

where $\eta_0, \phi_{10}$ and $\phi_{20}$ are given in terms of the
$\beta_i$ constants as
\begin{equation}
\eta_0=e^{\frac{\beta_1+\beta_2-6\beta_3}{12}}, \quad
\phi_{10}=\frac{3\beta_1+\beta_2-6\beta_3}{2\lambda_1},\quad
\phi_{20}=\frac{\beta_1+3\beta_2-6\beta_3}{2\lambda_2}.\label{phi2}
\end{equation}
 and the constants $a^a_{_\ell}=\frac{q^a_{_\ell}+3d^a_{_\ell}}{6q^a_{_\ell}}$,
$a^b_{_\ell}=\frac{q^b_{_\ell}+3d^b_{_\ell}}{6q^b_{_\ell}}$,
$\theta_1=\frac{3q^a_{_\ell}+3d^a_{_\ell}}{\lambda_1q^a_{_\ell}}$,
$\theta_2=\frac{q^b_{_\ell}+3d^b_{_\ell}}{\lambda_1 q^b_{_\ell}}$,
$\theta_3=\frac{q^a_{_\ell}+3d^a_{_\ell}}{\lambda_2\,q^a_{_\ell}}$,
$\theta_4=\frac{3
q^b_{_\ell}+3d^b_{_\ell}}{\lambda_2\,q^b_{_\ell}}$.

 Finally, the volume function and the scalar function
$\psi_i$ become
\begin{eqnarray}
\eta^3&=&\eta_0^3\,
Exp\left[-(q^b_{_\ell}d^{a2}_{_\ell}+q^a_{_\ell}d^{b2}_{_\ell}+4c_{_\ell}q^a_{_\ell}q^b_{_\ell})\frac{3p_3}{4q^a_{_\ell}q^b_{_\ell}}\Delta
t\right] \, \left[Cosh(r_1t-q_1)\right]^{3a^a_{_\ell}}
\left[Cosh(r_2t-q_2)\right]^{3a^b_{_\ell}},\label{eta_B}\\
\psi_1&=& \psi_{10} \,Exp\left[-\frac{3p_3}{2\lambda_1 \,
q^a_{_\ell}q^b_{_\ell}}\left(4c_{_\ell}q^a_{_\ell}q^b_{_\ell}+q^b_{_\ell}d^{a2}_{_\ell}
+q^a_{_\ell}d^{b2}_{_\ell} \right)\Delta t\right]
\,\left[Cosh^{\theta_1}\left(r_1t-q_1 \right) \, Cosh^{\theta_2}\left(r_2t-q_2 \right)\right]\label{psi1_B}\\
\psi_2&=& \psi_{20}\,Exp\left[- \frac{3p_3}{2\lambda_2 \,
q^a_{_\ell}q^b_{_\ell}}\left[4c_{_\ell}q^a_{_\ell}q^b_{_\ell}+d^{a2}_{_\ell}q^b_{_\ell}+
d^{b2}_{_\ell}q^a_{_\ell} \right]\Delta t \right]
\left[Cosh^{\theta_3}\left(r_1t-q_1 \right) \,
Cosh^{\theta_4}\left(r_2t-q_2 \right)\right] \label{psi2_B}
\end{eqnarray}
\subsubsection{Structure C}
For the last case, the Hamiltonian density takes the form
\begin{equation}
\rm {\cal H}= -q^a_{_\ell}\pi_1^2 +q^b_{_\ell}\pi_2^2
+c_{_\ell}\pi_3^2 +\left(d^a_{_\ell}\pi_1+ d^b_{_\ell}\pi_2 \right)
\pi_3-6\left(P_1^2+P_2^2+P_3^2\right)-24V_1e^{-\xi_1}-24V_2
e^{-\xi_2}, \label{hami-phantom-menos}
\end{equation}
the corresponding Hamilton equation become

\begin{eqnarray}\label{quintom-menos}
\rm \dot \xi_1&=& \rm -2q^a_{_\ell}\pi_1
+d^a_{_\ell}\pi_3, \nonumber\\
\rm \dot \xi_2&=& \rm 2q^b_{_\ell}\pi_2
+d^b_{_\ell}\pi_3, \\
\rm \dot \xi_3&=& \rm 2 c_{_\ell}\pi_3+
d^a_{_\ell}\pi_1+d^b_{_\ell}\pi_2  , \nonumber
\end{eqnarray}
the equations for $\rm \dot \pi_i$ remain the same as in
(\ref{ecs-mov-2}).

Taking the derivative of the first equation of (\ref{quintom-menos})
yields
\begin{equation}
\rm \ddot \xi_1= 48 V_1 q^a_{_\ell} e^{-\xi_1},
\end{equation}
which has a solution of the form
\begin{equation}
\rm e^{-\xi_1}=\frac{ r_1^2}{24 V_1 q^a_{_\ell}} \, Sech^2\left(r_1
t-q_1\right). \label{sol-quintom-menos1}
\end{equation}
and taking the derivative of the second equation of
(\ref{quintom-menos}) yields
\begin{equation}
\rm \ddot \xi_2= -48 V_2 q^a_{_\ell} e^{-\xi_1},
\end{equation}
which has a solution of the form
\begin{equation}
\rm e^{-\xi_2}=\frac{ r_2^2}{24 V_2 q^b_{_\ell}} \, Csch^2\left(r_2
t-q_2\right). \label{sol-quintom-menos2}
\end{equation}
Reinserting these solutions into Hamilton equations for the momenta,
we obtain
\begin{eqnarray}
\rm \pi_1 &=& \rm a_1 - \frac{r_1}{q^a_{_\ell}} \, Tanh\left(r_1
t-q_1
\right), \label{solution-p1}\\
\rm \pi_2 &=& \rm a_2 + \frac{r_2}{q^b_{_\ell}} \, Coth\left(r_2
t-q_2 \right). \label{solucion-p2}
\end{eqnarray}
When we include
(\ref{sol-quintom-menos1},\ref{sol-quintom-menos2},\ref{solution-p1},\ref{solucion-p2})
into the Hamiltonian constraint we obtain the constants
\begin{equation}
a_1=\frac{dl^a_{_\ell}}{2q^a_{_\ell}}p_3,\quad
a_2=-\frac{dl^b_{_\ell}}{2q^b_{_\ell}}p_3, \quad
p_3^2=4\frac{6q^a_{_\ell}q^b_{_\ell} n^2 +q^b_{_\ell}r_1^2 -
q^a_{_\ell}
r_2^2}{4c_{_\ell}q^a_{_\ell}q^b_{_\ell}+q^b_{_\ell}d^{a2}_{_\ell}-q^a_{_\ell}d^{b2}_{_\ell}},
\end{equation}
Now we are in position to write the solutions for the $\rm \xi_i$
coordinates, which read
\begin{eqnarray}
\rm \xi_1 &=& \rm \beta_1 +Ln\left[ Cosh^2\left(r_1 t -q_1 \right)\right],  \label{xi1-quintom-menos}\\
\xi_2 &=& \rm \beta_2+Ln\left[ Sinh^2\left(r_2 t -q_2 \right)\right], \label{xi2-quintom-menos}\\
\rm \xi_3&=&\rm \beta_3 +
\frac{p_3}{2q^a_{_\ell}q^b_{_\ell}}\left[4c_{_\ell}q^a_{_\ell}q^b_{_\ell}+q^b_{_\ell}d^{a2}_{_\ell}-q^a_{_\ell}d^{b2}_{_\ell}\right]\Delta
t-\frac{d^a_{_\ell}}{q^a_{_\ell}}\, Ln\,\left[Cosh\left(r_1t-q_1
\right) \right]+\frac{d^b_{_\ell}}{q^b_{_\ell}}\,
Ln\,\left[Sinh\left(r_2 t-q_2 \right) \right],
\label{xi3-quintom-menos}
\end{eqnarray}
 where the constants $\rm \beta_i\,(i=1,2,3)$ arise as
 integration constants. Applying the inverse canonical
transformation we obtain the solutions in the original variables
$\rm (\eta, \phi_1, \phi_2)$ as
\begin{eqnarray}
\eta&=&e^u=\eta_0\,
Exp\left[-\frac{p_3}{4q^a_{_\ell}q^b_{_\ell}}(4c_{_\ell}q^a_{_\ell}q^b_{_\ell}+q^b_{_\ell}d^{2a}_{\ell}-q^a_{_\ell}d^{b2}_{_\ell})\Delta
t\right] \, \left[Cosh(r_1t-q_1)\right]^{a^a_{_\ell}}
\left[Sinh(r_2t-q_2)\right]^{a^b_{_\ell}},\label{eta-quintom-menos}\\
\phi_1&=& \phi_{10} -\frac{3p_3}{2\lambda_1 \,
q^a_{_\ell}q^b_{_\ell}}\left(4c_{_\ell}q^a_{_\ell}q^b_{_\ell}+q^b_{_\ell}d^{a2}_{_\ell}
-q^a_{_\ell}d^{b2}_{_\ell} \right)\Delta t
+Ln\left[Cosh^{\theta_1}\left(r_1t-q_1 \right) \, Sinh^{\theta_2}\left(r_2t-q_2 \right)\right]\\
\phi_2&=& \phi_{20} - \frac{3p_3}{2\lambda_2 \,
q^a_{_\ell}q^b_{_\ell}}\left[4c_{_\ell}q^a_{_\ell}q^b_{_\ell}+d^{a2}_{_\ell}q^b_{_\ell}-
d^{b2}_{_\ell}q^a_{_\ell} \right]\Delta t
+Ln\left[Cosh^{\theta_3}\left(r_1t-q_1 \right) \,
Sinh^{\theta_4}\left(r_2t-q_2 \right)\right] \label{sols-pha-fin}
\end{eqnarray}
here $\eta_0, \phi_{10}$ and $\phi_{20}$ are given in terms of the
$\beta_i$ constants as
\begin{equation}
\eta_0=e^{\frac{\beta_1+\beta_2-6\beta_3}{12}}, \quad
\phi_{10}=\frac{3\beta_1+\beta_2-6\beta_3}{2\lambda_1},\quad
\phi_{20}=\frac{\beta_1+3\beta_2-6\beta_3}{2\lambda_2}.\label{phi2}
\end{equation}
 and the constants $a^a_{_\ell}=\frac{q^a_{_\ell}+3d^a_{_\ell}}{6q^a_{_\ell}}$,
$a^b_{_\ell}=\frac{q^b_{_\ell}-3d^b_{_\ell}}{6q^b_{_\ell}}$,
$\theta_1=\frac{3(q^a_{_\ell}+d^a_{_\ell})}{\lambda_1q^a_{_\ell}}$,
$\theta_2=\frac{q^b_{_\ell}-3d^b_{_\ell}}{\lambda_1q^b_{_\ell}}$,
$\theta_3=\frac{q^a_{_\ell}+3d^a_{_\ell}}{\lambda_2\,q^a_{_\ell}}$,
$\theta_4=\frac{3(
q^b_{_\ell}-d^b_{_\ell})}{\lambda_2\,q^b_{_\ell}}$.

The volume function and the scalar fields $\psi_i$ are
\begin{eqnarray}
\eta^3&=&\eta^3_0\,
Exp\left[-\frac{3p_3}{4q^a_{_\ell}q^b_{_\ell}}(4c_{_\ell}q^a_{_\ell}q^b_{_\ell}+q^b_{_\ell}d^{2a}_{\ell}-q^a_{_\ell}d^{b2}_{_\ell})\Delta
t\right] \, \left[Cosh(r_1t-q_1)\right]^{3a^a_{_\ell}}
\left[Sinh(r_2t-q_2)\right]^{3a^b_{_\ell}},\label{eta_C}\\
\psi_1&=& \psi_{10}\, Exp\left[ -\frac{3p_3}{2\lambda_1 \,
q^a_{_\ell}q^b_{_\ell}}\left(4c_{_\ell}q^a_{_\ell}q^b_{_\ell}+q^b_{_\ell}d^{a2}_{_\ell}
-q^a_{_\ell}d^{b2}_{_\ell} \right)\Delta t\right]
\left[Cosh^{\theta_1}\left(r_1t-q_1 \right) \, Sinh^{\theta_2}\left(r_2t-q_2 \right)\right]\label{psi1_C}\\
\psi_2&=& \psi_{20} \, Exp\left[ - \frac{3p_3}{2\lambda_2 \,
q^a_{_\ell}q^b_{_\ell}}\left[4c_{_\ell}q^a_{_\ell}q^b_{_\ell}+d^{a2}_{_\ell}q^b_{_\ell}-
d^{b2}_{_\ell}q^a_{_\ell} \right]\Delta t\right]
\left[Cosh^{\theta_3}\left(r_1t-q_1 \right) \,
Sinh^{\theta_4}\left(r_2t-q_2 \right)\right]
\label{psi2_C}
\end{eqnarray}
On the other hand, in the plots presented in the following sections, the horizontal axis represents the evolution parameter introduced after fixing the lapse function, $N=24e^{3u}$, in the Hamiltonian formulation. Therefore, the variable $t$ is not the physical cosmic time but the time parameter associated with the chosen gauge. The interval $0\le t\le 0.5$ describes the expanding (future) branch of the exact cosmological solutions, starting from the reference instant $t=0$.

The vertical axis displays the normalized evolution of the average volume function $\eta^3(t)$ together with the scalar fields $\psi_1(t)$ and $\psi_2(t)$. The average volume can be written as
\[
\eta^3(t)=\eta_0^3\,\mathcal{F}(t),
\]
where $\eta_0^3$ is an integration constant that fixes the reference (initial) volume and $\mathcal{F}(t)$ is a dimensionless function. It follows that the plotted values should be interpreted as the evolution of the cosmological volume relative to the reference volume $\eta_0^3$, rather than as an absolute physical volume. Likewise, the scalar fields satisfy
\[
\psi_1(t)=\psi_{10}\,\mathcal{G}_1(t),\qquad
\psi_2(t)=\psi_{20}\,\mathcal{G}_2(t),
\]
where $\psi_{10}$ and $\psi_{20}$ determine the normalization of each field, while the dimensionless functions $\mathcal{G}_1(t)$ and $\mathcal{G}_2(t)$ describe their time evolution. While the amplitudes displayed in the plots depend on the selected integration constants, the functional behavior is completely determined by the underlying analytical solutions.

For the parameter choice $\lambda_1=2$, $\lambda_2=4$, and $\ell=9/16$, the figure shows that the average cosmological volume increases monotonically during the future evolution, remaining between the two scalar-field solutions. 

The constants $\eta_0^3$, $\psi_{10}$ and $\psi_{20}$ are integration
constants that set the normalization scale of the average volume and of the
two scalar fields, respectively. In particular, $\eta_0^3$ fixes the reference
scale of the comoving volume $V(t)=A(t)B(t)C(t)=\eta^3(t)$, while
$\psi_{10}$ and $\psi_{20}$ determine the corresponding reference amplitudes
of the scalar fields $\psi_1$ and $\psi_2$. Since these constants only rescale
the solutions, the quantities plotted in the figures are normalized as
$\eta^3/\eta_0^3$, $\psi_1/\psi_{10}$ and $\psi_2/\psi_{20}$, which display
the relative evolution of the volume and scalar fields independently of the
choice of normalization.

\section{Quintessence multiscalar fields, $m^{11}=m^{22}=+1$}

For the quintessence sector we set $m^{11}=m^{22}=+1$. In this case,
the parameters $\lambda_i$ are positive real numbers associated with
the powers of the original scalar fields $\psi_i$. The constraint on
the mixed kinetic coefficient gives two possible branches,
\begin{eqnarray}
m^{12}_{+}&=&\frac{\lambda_1 \lambda_2}{6}\left(1+ \sqrt{1+
\frac{36}{\lambda_1^2 \lambda_2^2}}\right),
\label{positive}\\
m^{12}_{-}&=&\frac{\lambda_1 \lambda_2}{6}\left( 1-\sqrt{1+
\frac{36}{\lambda_1^2 \lambda_2^2}}\right). \label{negative}
\end{eqnarray}
It follows that $m^{12}_{+}>m^{12}_{-}$, with $m^{12}_{-}<0$.
Introducing the dimensionless parameter
\begin{equation}
\ell=\frac{36}{\lambda_1^2\lambda_2^2}>0,
\end{equation}
the two branches can be written as
\begin{equation}
m^{12}_{+}=
\frac{\lambda_1\lambda_2}{6}\left(1+\sqrt{1+\ell}\right)>0,
\qquad
m^{12}_{-}=-
\frac{\lambda_1\lambda_2}{6}\left(\sqrt{1+\ell}-1\right)<0.
\end{equation}
The corresponding values of the determinant parameter are
\begin{equation}
\triangle_{+}=
-\frac{2}{\ell}\left(1+\sqrt{1+\ell}\right)<0,
\qquad
\triangle_{-}=
\frac{2}{\ell}\left(\sqrt{1+\ell}-1\right)>0.
\end{equation}


\subsubsection{$m^{12}_-$ case}

For the negative branch,
\begin{equation}
m^{12}_{-}=-
\frac{1}{6}\left(\sqrt{1+\ell}-1\right)\lambda_1\lambda_2,
\end{equation}
the Hamiltonian density becomes
\begin{equation}
{\cal H}= -q^a_{_\ell}\pi_1^2 -q^b_{_\ell}\pi_2^2
+c_{_\ell}\pi_3^2 +\left(d^a_{_\ell}\pi_1+ d^b_{_\ell}\pi_2 \right)
\pi_3-6\left(P_1^2+P_2^2+P_3^2\right)-24V_1e^{-\xi_1}-24V_2
e^{-\xi_2}, \label{hami-bianchi-q}
\end{equation}
which corresponds to Hamiltonian structure A. Therefore, for a given
set of model parameters, the exact solutions are determined by
Eqs.~\eqref{eta_A}, \eqref{psi1_A}, and \eqref{psi2_A}, with
\begin{eqnarray}
q^a_{_\ell}&=&\frac{6\ell
\lambda_1^2-36\left(\sqrt{1+\ell}-1\right)}{\sqrt{\ell+1}-1},
\nonumber\\
q^b_{_\ell}&=&\frac{6\ell
\lambda_2^2-36\left(\sqrt{1+\ell}-1\right)}{\sqrt{\ell+1}-1},
\nonumber\\
c_{_\ell}&=&\frac{14\left(\sqrt{1+\ell}-1\right)-\frac{\ell}{6}
\left(\lambda_1^2+\lambda_2^2\right)}{\sqrt{\ell+1}-1},
\nonumber\\
d^a_{_\ell}&=&\frac{36 \left(\sqrt{1+\ell}-1\right)-2\ell
\lambda_1^2}{\sqrt{\ell+1}-1}, \nonumber\\
d^b_{_\ell}&=&\frac{36 \left(\sqrt{1+\ell}-1\right)-2\ell
\lambda_2^2}{\sqrt{\ell+1}-1}.
\label{q-menos}
\end{eqnarray}

\begin{figure}[ht!]
\begin{center}
\captionsetup{width=.8\textwidth}
\includegraphics[scale=0.5]{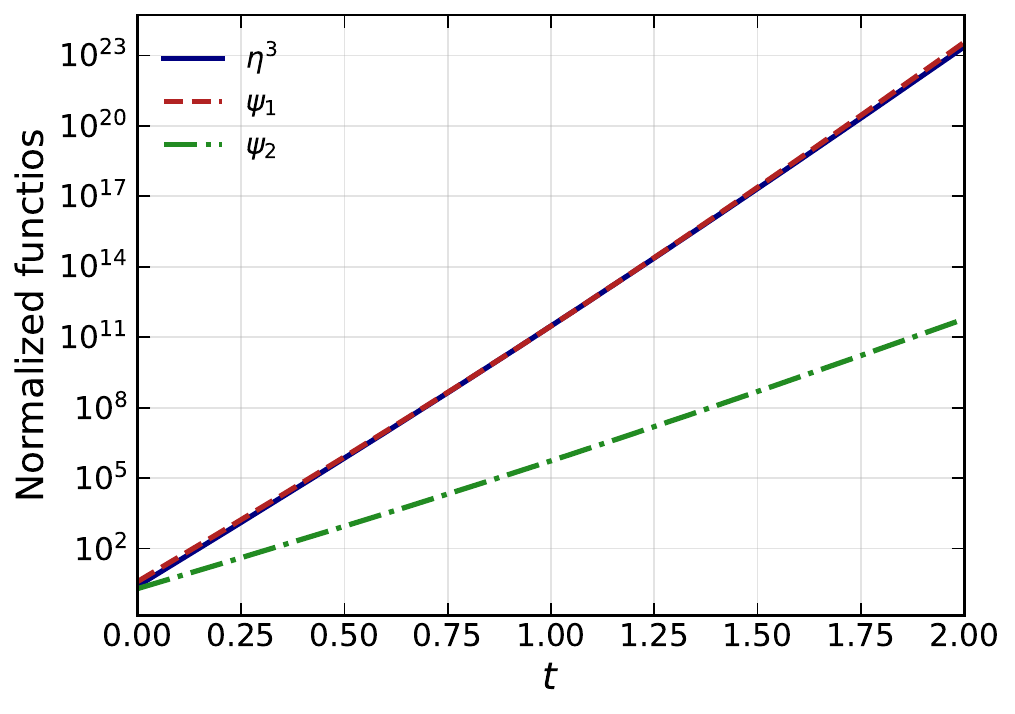}
\includegraphics[scale=0.5]{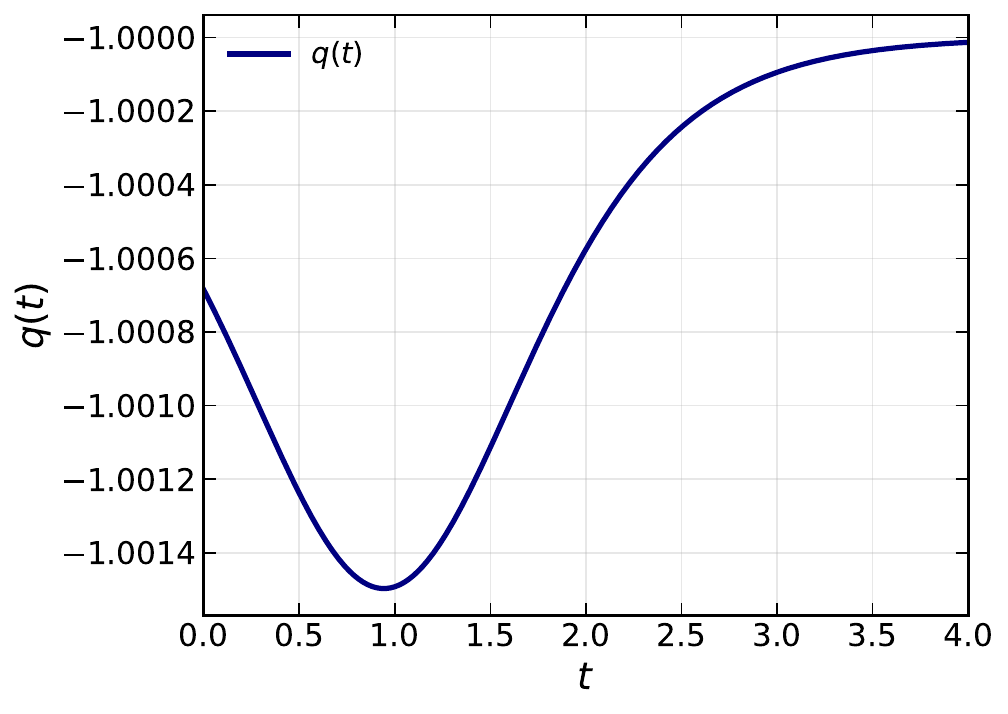}
\caption{
Evolution of the normalized average volume function,
$\eta^{3}/\eta_{0}^{3}$, together with the normalized scalar fields
$\psi_{1}/\psi_{10}$ and $\psi_{2}/\psi_{20}$ (left panel), and the
corresponding deceleration parameter $q(t)$ (right panel) for the
Quintessence$-$ cosmological branch with inverse power-law potential
$V(\psi_{1},\psi_{2})=\psi_{1}^{-2}+\psi_{2}^{-4}$. The parameters are
chosen as $\lambda_{1}=2$, $\lambda_{2}=4$, and $\ell=9/16$, with
$r_{1}=r_{2}=q_{1}=q_{2}=1$, while the remaining constants are fixed by
the exact solution. The average volume and scalar fields evolve
monotonically, whereas the deceleration parameter rapidly approaches
the de Sitter value $q=-1$, indicating that the expansion remains
accelerated throughout the evolution.
}
\label{Qumenos}
\end{center}
\end{figure}

The exact solution associated with the Quintessence$-$ branch describes
an almost de Sitter cosmological evolution from the beginning of the
expansion. As shown in Fig.~\ref{Qumenos}, the normalized average
volume function and both scalar fields evolve monotonically, although
the two fields contribute with different amplitudes. While $\psi_1$
closely follows the growth of the average volume, $\psi_2$ remains
subdominant throughout the interval considered. The reconstructed
deceleration parameter stays extremely close to $q=-1$, exhibiting only
a small transient departure before converging to the asymptotic de
Sitter value. Consequently, this branch represents the solution whose
average dynamics are closest to an exponentially expanding universe
throughout the interval considered.

\subsubsection{$m^{12}_+$ case}

For the positive branch,
\begin{equation}
m^{12}_{+}=
\frac{1}{6}\left(1+\sqrt{1+\ell}\right)\lambda_1\lambda_2,
\end{equation}
the Hamiltonian density is
\begin{equation}
{\cal H}= q^a_{_\ell} \pi_1^2 +q^b_{_\ell}\pi_2^2
+c_{_\ell}\pi_3^2 +\left(d^a_{_\ell}\pi_1+d^b_{_\ell}\pi_2 \right)
\pi_3-6\left(P_1^2+P_2^2+P_3^2\right)-24V_1e^{-\xi_1}-24V_2
e^{-\xi_2}, \label{hami-quinte-mas}
\end{equation}
which is described by Hamiltonian structure B. The exact solutions are
therefore obtained from Eqs.~\eqref{eta_B}, \eqref{psi1_B}, and
\eqref{psi2_B} after specifying
\begin{eqnarray}
q^a_{_\ell}&=&\frac{6\ell
\lambda_1^2+36\left(1+\sqrt{1+\ell}\right)}{1+\sqrt{1+\ell}},
\nonumber\\
q^b_{_\ell}&=&\frac{6\ell
\lambda_2^2+36\left(1+\sqrt{1+\ell}\right)}{1+\sqrt{1+\ell}},
\nonumber\\
c_{_\ell}&=&\frac{14\left(1+\sqrt{1+\ell}\right)+\frac{\ell}{6}
\left(\lambda_1^2+\lambda_2^2\right)}{1+\sqrt{1+\ell}},
\nonumber\\
d^a_{_\ell}&=&\frac{36 \left(1+\sqrt{1+\ell}\right)+2\ell
\lambda_1^2}{1+\sqrt{1+\ell}}, \nonumber\\
d^b_{_\ell}&=&\frac{36 \left(1+\sqrt{1+\ell}\right)+2\ell
\lambda_2^2}{1+\sqrt{1+\ell}}.
\label{q-mas}
\end{eqnarray}

\begin{figure}[ht!]
\begin{center}
\captionsetup{width=.8\textwidth}
\includegraphics[scale=0.5]{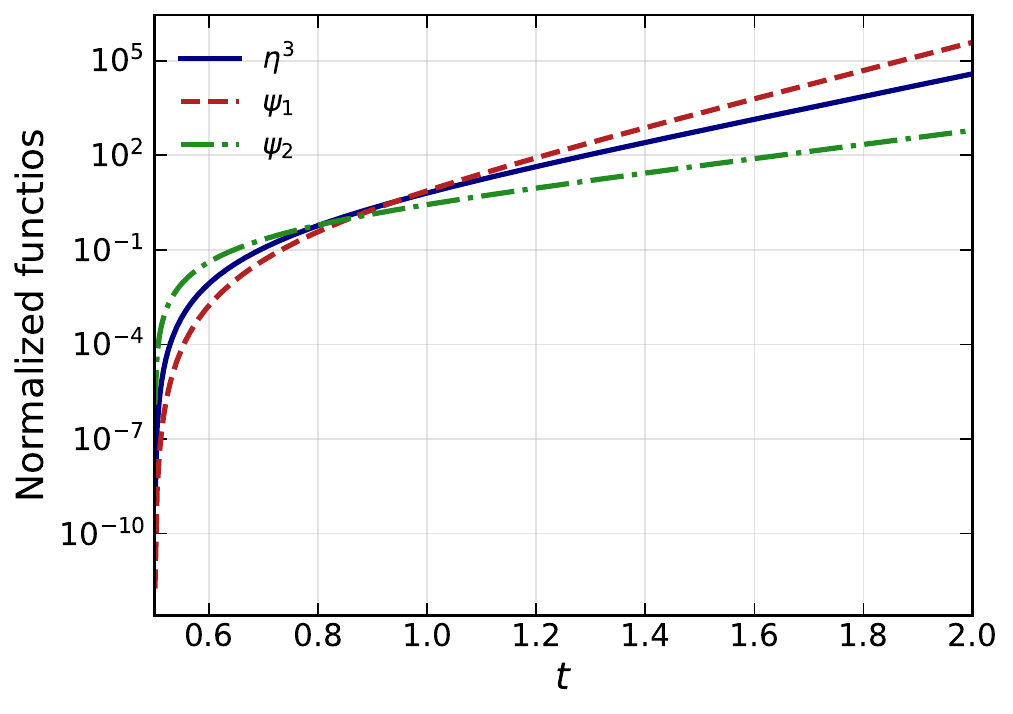}
\includegraphics[scale=0.5]{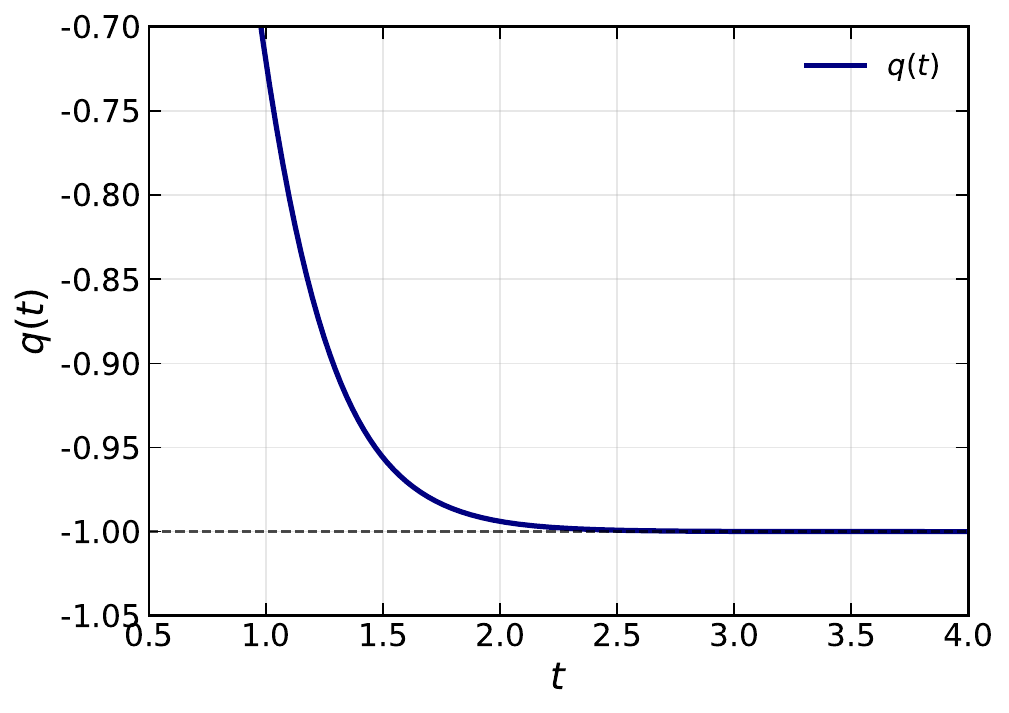}
\caption{
Comparison between the normalized average volume function
$\eta^{3}/\eta_{0}^{3}$ and the normalized scalar fields
$\psi_{1}/\psi_{10}$ and $\psi_{2}/\psi_{20}$ (left panel), together
with the corresponding deceleration parameter $q(t)$ (right panel) for
the Quintessence$+$ cosmological branch. The model parameters are
chosen as $\lambda_{1}=2$, $\lambda_{2}=4$, $\ell=9/16$, with
$r_{1}=r_{2}=2$ and $q_{1}=q_{2}=1$, while the remaining constants are
fixed by the exact analytical solution. In contrast to the
Quintessence$-$ branch, the scalar fields display noticeably different
growth rates relative to the average volume function. Nevertheless, the
cosmological expansion rapidly settles into an accelerated regime, with
$q(t)$ smoothly approaching the de Sitter limit $q=-1$ from the
quintessence side.
}
\label{Qumas_all}
\end{center}
\end{figure}

Unlike the previous branch, the Quintessence$+$ solution exhibits a
more pronounced hierarchy between the scalar fields while preserving
the same accelerated cosmological behavior. Figure~\ref{Qumas_all}
shows that $\psi_1$ eventually grows slightly faster than the average
volume function, whereas $\psi_2$ remains subdominant throughout the
evolution. Despite these differences in the scalar-field dynamics, the
reconstructed deceleration parameter remains negative and rapidly
approaches the de Sitter limit. Thus, both quintessence branches share
the same late-time attractor, while differing in their transient
scalar-field evolution.

\newpage

\section{Phantom multiscalar fields, $m^{11}=m^{22}=-1$}

For the phantom sector we set $m^{11}=m^{22}=-1$. The constraint on the
mixed coefficient $m^{12}$ has the same formal structure as in the
quintessence case, and therefore both branches $m^{12}_{\pm}$ must be
analyzed separately.

\subsubsection{$m^{12}_-$ case}

For
\begin{equation}
m^{12}_{-}=-
\frac{1}{6}\left(\sqrt{1+\ell}-1\right)\lambda_1\lambda_2,
\end{equation}
the Hamiltonian density becomes
\begin{equation}
{\cal H}= q^a_{_\ell} \pi_1^2 +q^b_{_\ell}\pi_2^2
+c_{_\ell}\pi_3^2 +\left(d^a_{_\ell}\pi_1+d^b_{_\ell}\pi_2 \right)
\pi_3-6\left(P_1^2+P_2^2+P_3^2\right)-24V_1e^{-\xi_1}-24V_2
e^{-\xi_2}, \label{hami-phantom-menos}
\end{equation}
which corresponds to Hamiltonian structure B. Hence, the exact
solutions are obtained from Eqs.~\eqref{eta_B}, \eqref{psi1_B}, and
\eqref{psi2_B}, with
\begin{eqnarray}
q^a_{_\ell}&=&\frac{6\ell
\lambda_1^2+36\left(\sqrt{1+\ell}-1\right)}{\sqrt{1+\ell}-1},
\nonumber\\
q^b_{_\ell}&=&\frac{6\ell
\lambda_2^2+36\left(\sqrt{1+\ell}-1\right)}{\sqrt{1+\ell}-1},
\nonumber\\
c_{_\ell}&=&\frac{14\left(\sqrt{1+\ell}-1\right)+\frac{\ell}{6}
\left(\lambda_1^2+\lambda_2^2\right)}{\sqrt{1+\ell}-1},
\nonumber\\
d^a_{_\ell}&=&\frac{36 \left(\sqrt{1+\ell}-1\right)+2\ell
\lambda_1^2}{\sqrt{1+\ell}-1}, \nonumber\\
d^b_{_\ell}&=&\frac{36 \left(\sqrt{1+\ell}-1\right)+2\ell
\lambda_2^2}{\sqrt{1+\ell}-1}.
\label{phantom-menos}
\end{eqnarray}

\begin{figure}[ht!]
\begin{center}
\captionsetup{width=.8\textwidth}
\includegraphics[scale=0.5]{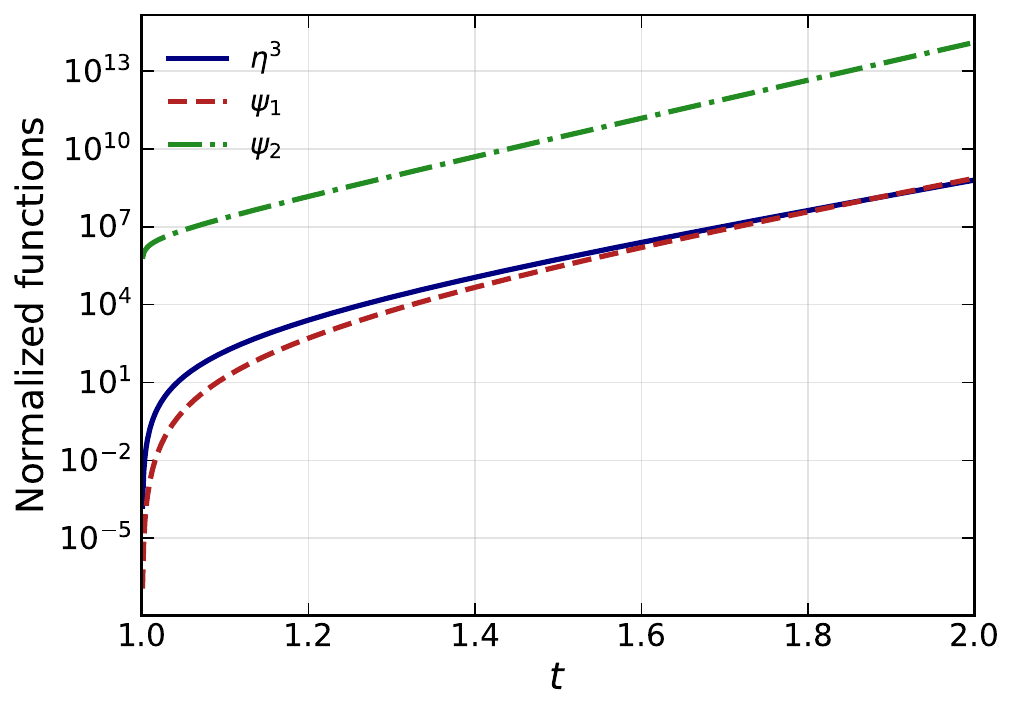}
\includegraphics[scale=0.5]{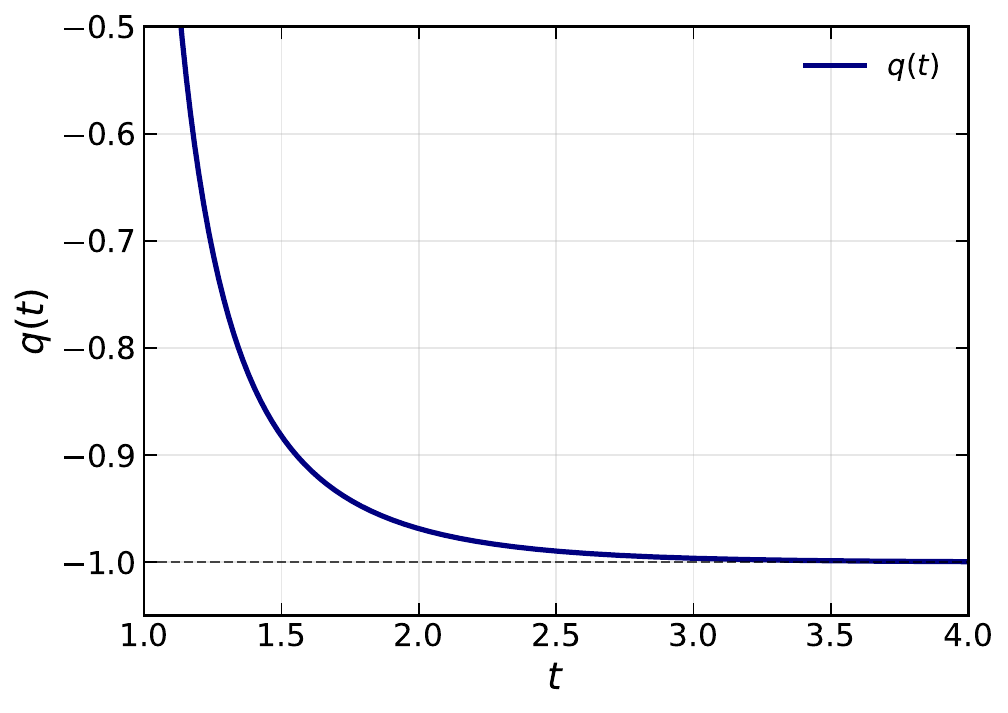}
\caption{
The Phantom $m^{12}_-$ cosmological branch exhibits a markedly different
hierarchy among the dynamical variables. The left panel compares the
normalized average volume function $\eta^{3}/\eta_{0}^{3}$ with the
normalized scalar fields $\psi_{1}/\psi_{10}$ and
$\psi_{2}/\psi_{20}$, whereas the right panel shows the corresponding
deceleration parameter $q(t)$. The parameters are chosen as
$m_{2}^{-1}=1$, $\lambda_{1}=2$, $\lambda_{2}=4$,
$\ell=9/16$, and $n=r_{1}=r_{2}=q_{1}=q_{2}=1$, while the remaining
constants are determined by the exact analytical solution. In contrast
to the quintessence branches, the second scalar field becomes the
dominant contribution during the evolution, although the average
expansion remains regular. The deceleration parameter decreases
monotonically toward the de Sitter limit $q=-1$, confirming that the
late-time accelerated expansion is preserved despite the phantom
kinetic sector.
}
\label{volumen-phantom-menos}
\end{center}
\end{figure}

A distinctive feature of the Phantom$-$ branch is the dominant role
played by the second scalar field. As illustrated in
Fig.~\ref{volumen-phantom-menos}, $\psi_2$ rapidly exceeds both the
normalized average volume and $\psi_1$, becoming the leading scalar
contribution during the cosmological evolution. Nevertheless, the
average expansion remains regular and accelerated, with the
deceleration parameter monotonically approaching the de Sitter limit.
Thus, the phantom kinetic sector substantially modifies the transient
scalar-field hierarchy without altering the universal asymptotic
behavior of the average cosmological expansion.

\subsubsection{$m^{12}_+$ case}

For
\begin{equation}
m^{12}_{+}=
\frac{1}{6}\left(\sqrt{1+\ell}+1\right)\lambda_1\lambda_2,
\end{equation}
the Hamiltonian density is rewritten as
\begin{equation}
{\cal H}= q^a_{_\ell}\pi_1^2 +q^b_{_\ell}\pi_2^2
+c_{_\ell}\pi_3^2 +\left(d^a_{_\ell}\pi_1+ d^b_{_\ell}\pi_2 \right)
\pi_3-6\left(P_1^2+P_2^2+P_3^2\right)-24V_1e^{-\xi_1}-24V_2
e^{-\xi_2}, \label{hami-pha-mas}
\end{equation}
which is again described by Hamiltonian structure B. The exact
solutions follow from Eqs.~\eqref{eta_B}, \eqref{psi1_B}, and
\eqref{psi2_B}, after specifying
\begin{eqnarray}
q^a_{_\ell}&=&\frac{36\left(\sqrt{1+\ell}+1\right)-6\ell
\lambda_1^2}{\sqrt{\ell+1}+1},
\nonumber\\
q^b_{_\ell}&=&\frac{36\left(\sqrt{1+\ell}+1\right)-6\ell
\lambda_2^2}{\sqrt{\ell+1}+1},
\nonumber\\
c_{_\ell}&=&\frac{14\left(\sqrt{1+\ell}+1\right)-\frac{\ell}{6}
\left(\lambda_1^2+\lambda_2^2\right)}{\sqrt{\ell+1}+1},
\nonumber\\
d^a_{_\ell}&=&\frac{36 \left(\sqrt{1+\ell}+1\right)-2\ell
\lambda_1^2}{\sqrt{\ell+1}+1}, \nonumber\\
d^b_{_\ell}&=&\frac{36 \left(\sqrt{1+\ell}+1\right)-2\ell
\lambda_2^2}{\sqrt{\ell+1} +1}.
\label{phantom-mas}
\end{eqnarray}

\begin{figure}[ht!]
\begin{center}
\captionsetup{width=.8\textwidth}
\includegraphics[scale=0.5]{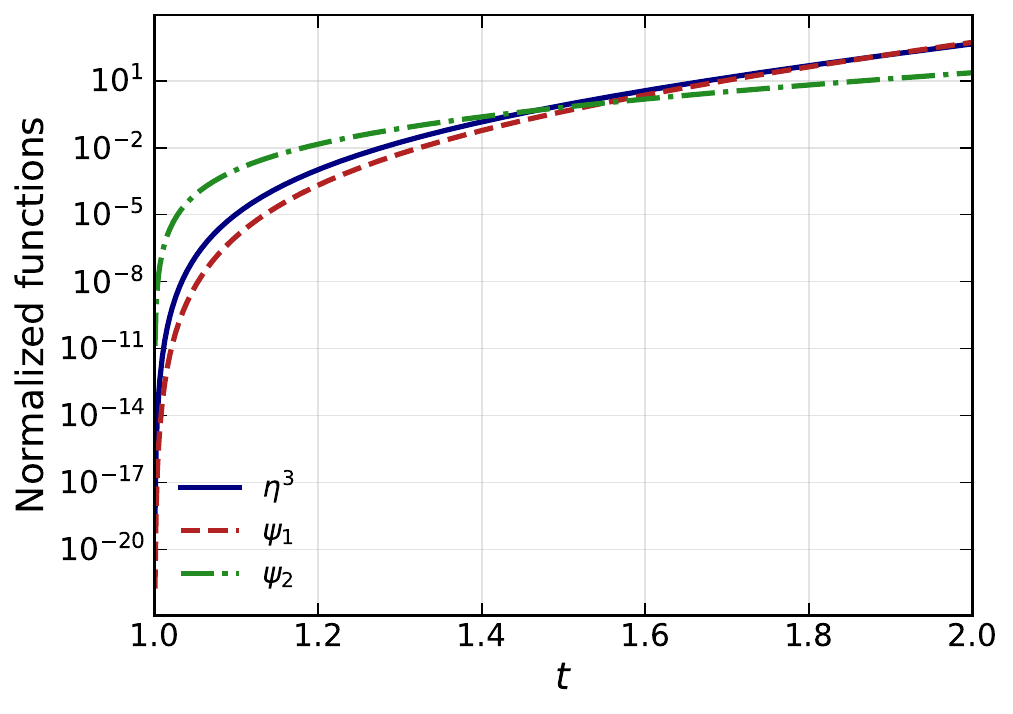}
\includegraphics[scale=0.5]{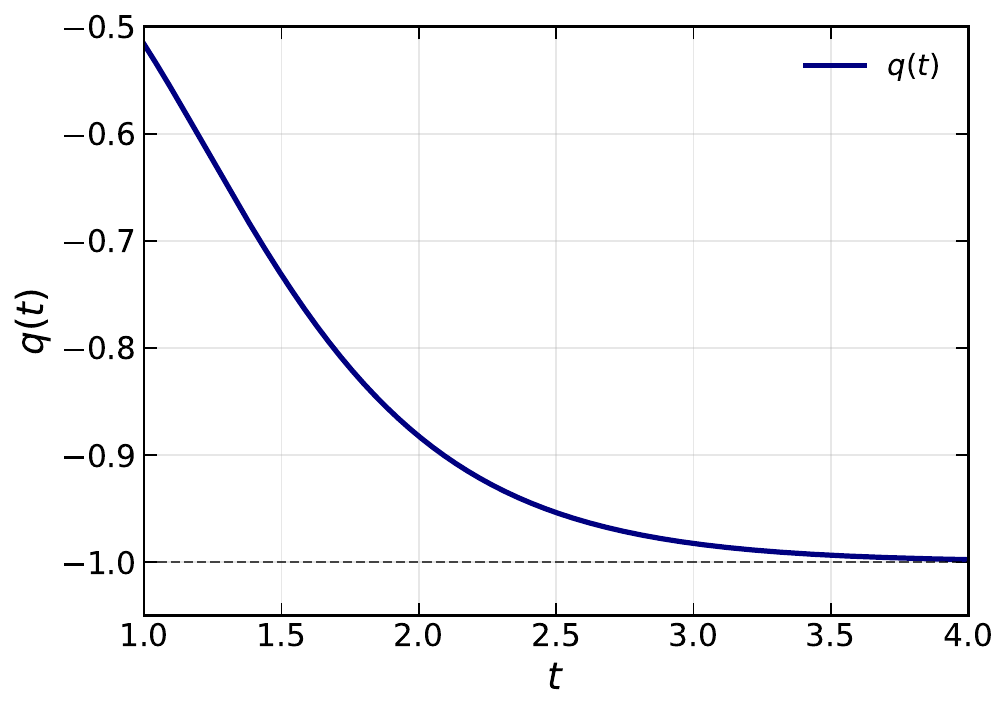}
\caption{
Despite belonging to the phantom sector, the Phantom $m^{12}_+$ cosmological
branch exhibits a smooth and regular cosmological evolution. The left
panel compares the normalized average volume function
$\eta^{3}/\eta_{0}^{3}$ with the normalized scalar fields
$\psi_{1}/\psi_{10}$ and $\psi_{2}/\psi_{20}$, whereas the right panel
shows the corresponding deceleration parameter $q(t)$. The parameters
are chosen as $\lambda_{1}=2$, $\lambda_{2}=4$, $\ell=9/16$, with
$r_{1}=r_{2}=q_{1}=q_{2}=1$, while the remaining constants are fixed by
the exact analytical solution. The average volume and the first scalar
field evolve with comparable growth rates, whereas the second scalar
field remains dynamically distinct throughout the evolution. At the
same time, the deceleration parameter decreases monotonically toward
$q=-1$, demonstrating that the accelerated expansion approaches the
same asymptotic de Sitter state found in the remaining exact branches.
}
\label{volumen-pha-mas}
\end{center}
\end{figure}

The Phantom$+$ branch displays a smoother and more balanced scalar-field
evolution than the previous phantom solution. As shown in
Fig.~\ref{volumen-pha-mas}, the average volume and the first scalar field
evolve with comparable growth rates, whereas $\psi_2$ remains
dynamically distinct. The deceleration parameter remains negative
throughout the physical interval and converges smoothly toward $q=-1$.
This shows that, despite the phantom character of the kinetic sector,
the solution remains dynamically regular and approaches the same
asymptotic de Sitter state as the other physically admissible branches.

\section{Quintom multiscalar fields, $m^{11}=+1, m^{22}=-1$}

For the mixed, or quintom, sector we take $m^{11}=+1$ and
$m^{22}=-1$. In this case, the constraint on the mixed kinetic
coefficient gives
\begin{eqnarray}
m^{12}_{+}&=&\frac{\lambda_1 \lambda_2}{6}\left(1+ \sqrt{1-
\ell}\right),
\label{qt-positive}\\
m^{12}_{-}&=&\frac{\lambda_1 \lambda_2}{6}\left(
1-\sqrt{1-\ell}\right). \label{qt-negative}
\end{eqnarray}
The two branches are real for the parameter region in which
$1-\ell\geq0$. We analyze them separately below.

\subsubsection{$m^{12}_-$ case}

For
\begin{equation}
m^{12}_{-}=
\frac{1}{6}\left(1-\sqrt{1-\ell}\right)\lambda_1\lambda_2,
\end{equation}
the Hamiltonian density becomes
\begin{equation}
{\cal H}= -q^a_{_\ell}\pi_1^2 +q^b_{_\ell}\pi_2^2
+c_{_\ell}\pi_3^2 +\left(d^a_{_\ell}\pi_1+ d^b_{_\ell}\pi_2 \right)
\pi_3-6\left(P_1^2+P_2^2+P_3^2\right)-24V_1e^{-\xi_1}-24V_2
e^{-\xi_2}, \label{hami-quintom-menos}
\end{equation}
corresponding to Hamiltonian structure C. Thus, for a given set of
parameters, the exact solutions are fully determined by
Eqs.~\eqref{eta_C}, \eqref{psi1_C}, and \eqref{psi2_C}, with
\begin{eqnarray}
q^a_{_\ell}&=&\frac{6\ell
\lambda_1^2-36\left(1-\sqrt{1-\ell}\right)}{1-\sqrt{1-\ell}},
\nonumber\\
q^b_{_\ell}&=&\frac{36\left(1-\sqrt{1-\ell}\right)+6\ell
\lambda_2^2}{1-\sqrt{1-\ell}},
\nonumber\\
c_{_\ell}&=&\frac{14\left(1-\sqrt{1-\ell}\right)+\frac{\ell}{6}
\left(\lambda_2^2-\lambda_1^2\right)}{1-\sqrt{1-\ell}},
\nonumber\\
d^a_{_\ell}&=&\frac{36 \left(1-\sqrt{1-\ell}\right)-2\ell
\lambda_1^2}{1-\sqrt{1-\ell}}, \nonumber\\
d^b_{_\ell}&=&\frac{36 \left(1-\sqrt{1-\ell}\right)+2\ell
\lambda_2^2}{1-\sqrt{1-\ell}}.
\end{eqnarray}

\begin{figure}[ht!]
\begin{center}
\captionsetup{width=.8\textwidth}
\includegraphics[scale=0.5]{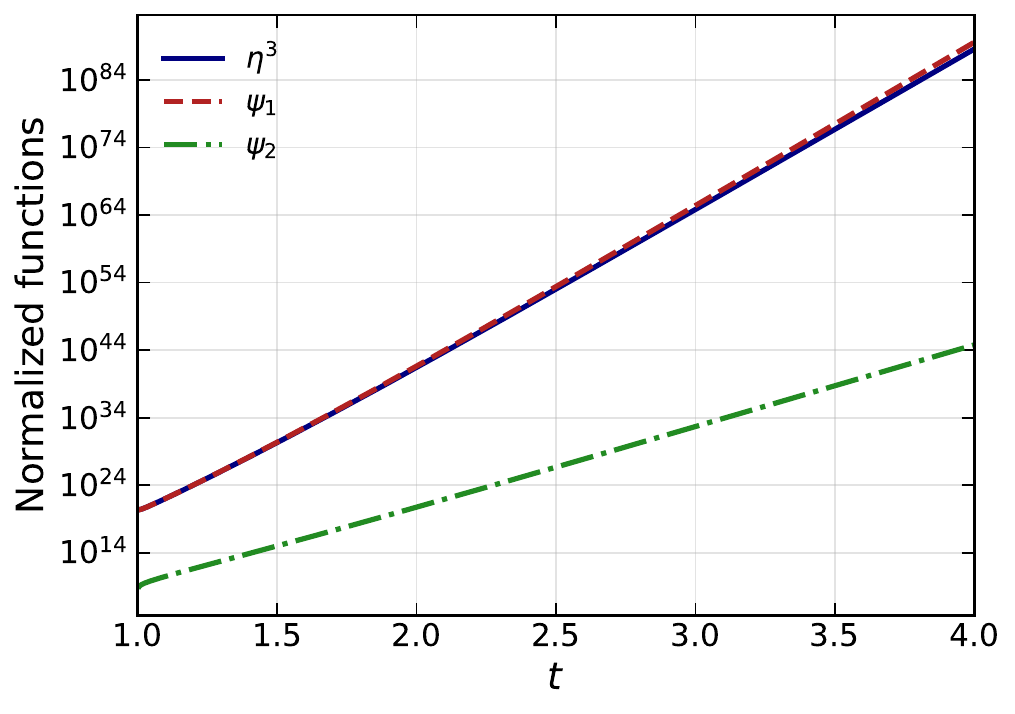}
\includegraphics[scale=0.5]{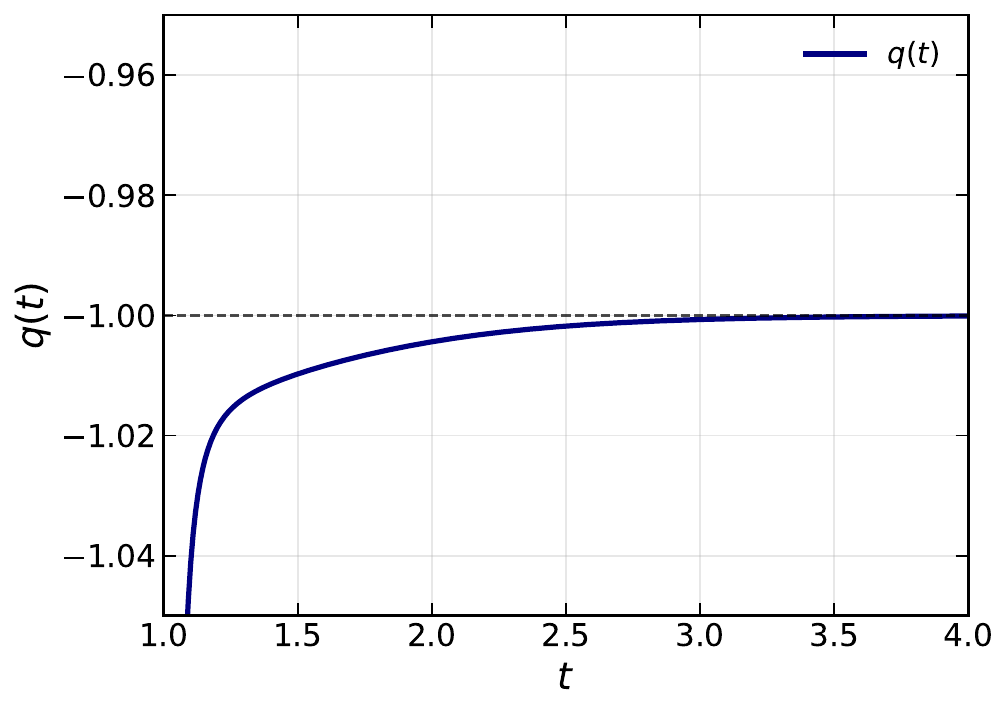}
\caption{
The Quintom $m^{12}_-$ branch combines canonical and phantom kinetic
contributions, giving rise to the richest dynamical evolution among the
exact solutions. The left panel shows the normalized average volume
function $\eta^{3}/\eta_{0}^{3}$ together with the normalized scalar
fields $\psi_{1}/\psi_{10}$ and $\psi_{2}/\psi_{20}$, while the right
panel displays the corresponding deceleration parameter $q(t)$. The
parameters are chosen as $\lambda_{1}=2$, $\lambda_{2}=4$,
$\ell=9/16$, with $r_{1}=r_{2}=q_{1}=q_{2}=1$, whereas the remaining
constants are fixed by the exact analytical solution. The second scalar
field rapidly dominates the evolution, illustrating the strong
interplay between the two kinetic sectors. In contrast to the previous
branches, the deceleration parameter approaches the de Sitter limit
$q=-1$ from the phantom regime ($q<-1$), revealing a transient
super-accelerated phase before reaching the universal late-time
attractor.
}
\label{volumen-quintom-menos}
\end{center}
\end{figure}

Among the six exact branches, the Quintom$-$ solution exhibits the
richest transient dynamics. Figure~\ref{volumen-quintom-menos} shows
that the second scalar field rapidly dominates the evolution, while the
reconstructed deceleration parameter approaches the de Sitter attractor
from the phantom regime, $q<-1$. This transient super-accelerated phase
is a characteristic signature of the mixed kinetic sector and
distinguishes this solution from the remaining branches, which approach
the attractor from the quintessence side. Nevertheless, the late-time
behavior is again governed by the same asymptotic de Sitter state.

\newpage
\subsubsection{$m^{12}_+$ case}

For
\begin{equation}
m^{12}_{+}=
\frac{1}{6}\left(1+\sqrt{1-\ell}\right)\lambda_1\lambda_2,
\end{equation}
the Hamiltonian density is
\begin{equation}
{\cal H}= q^a_{_\ell}\pi_1^2+q^b_{_\ell}\pi_2^2
+c_{_\ell}\pi_3^2 +\left(d^a_{_\ell}\pi_1+ d^b_{_\ell}\pi_2 \right)
\pi_3-6\left(P_1^2+P_2^2+P_3^2\right)-24V_1e^{-\xi_1}-24V_2
e^{-\xi_2}, \label{hami-quintom-mas}
\end{equation}
which corresponds to Hamiltonian structure B. The complete set of exact
solutions is obtained by specifying the parameters in
Eqs.~\eqref{eta_B}, \eqref{psi1_B}, and \eqref{psi2_B}:
\begin{eqnarray}
q^a_{_\ell}&=&\frac{36\left(1+\sqrt{1-\ell}\right)-6\ell
\lambda_1^2}{1+\sqrt{1-\ell}},
\nonumber\\
q^b_{_\ell}&=&\frac{6\ell
\lambda_2^2+36\left(1+\sqrt{1-\ell}\right)}{1+\sqrt{1-\ell}},
\nonumber\\
c_{_\ell}&=&\frac{14\left(1+\sqrt{1-\ell}\right)+\frac{\ell}{6}
\left(-\lambda_1^2+\lambda_2^2\right)}{1+\sqrt{1-\ell}},
\nonumber\\
d^a_{_\ell}&=&\frac{36 \left(1+\sqrt{1-\ell}\right)-2\ell
\lambda_1^2}{1+\sqrt{1-\ell}}, \nonumber\\
d^b_{_\ell}&=&\frac{36 \left(1+\sqrt{1-\ell}\right)+2\ell
\lambda_2^2}{1+\sqrt{1-\ell}}.
\label{quintom-mas}
\end{eqnarray}

\begin{figure}[h!]
\begin{center}
\captionsetup{width=.8\textwidth}
\includegraphics[scale=0.5]{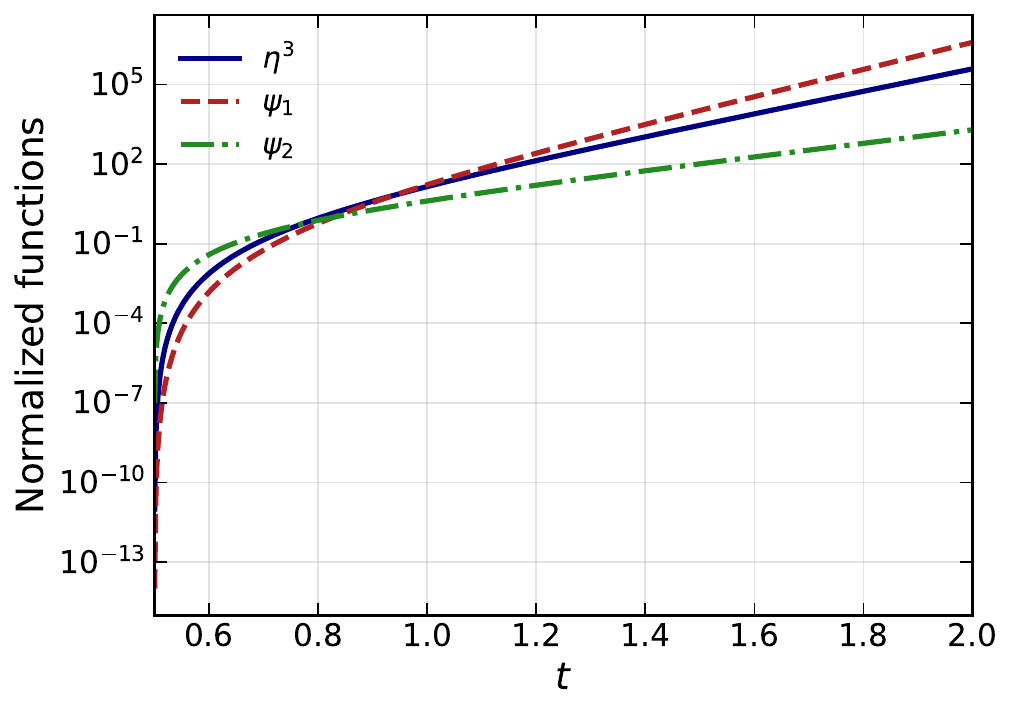}
\includegraphics[scale=0.5]{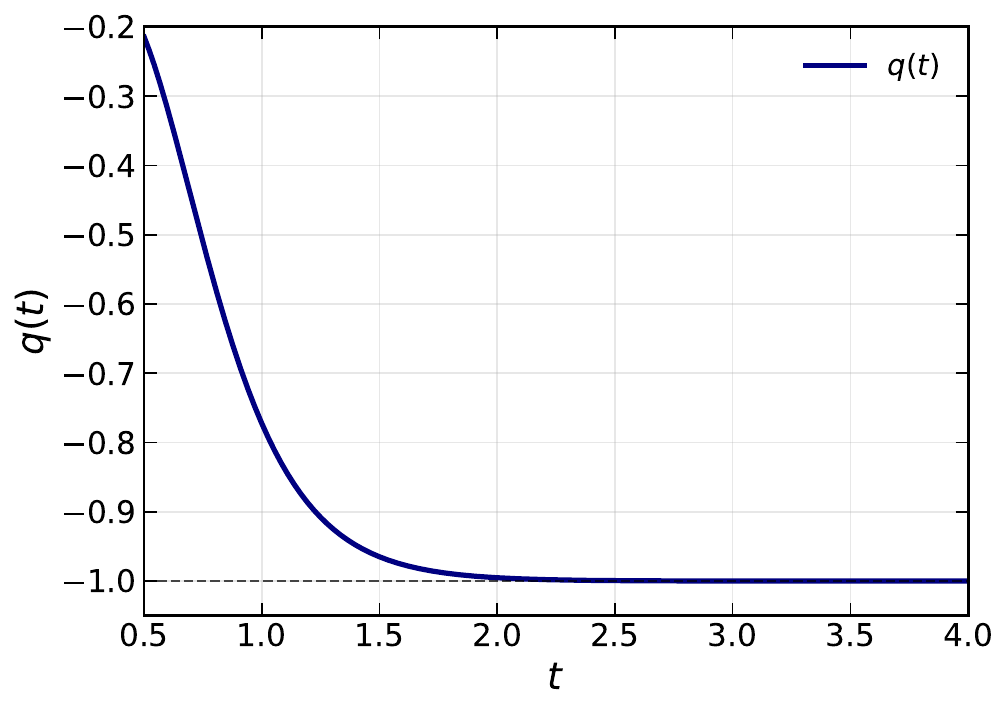}
\caption{
The Quintom$+$ branch provides a second realization of the mixed
kinetic sector, exhibiting a smoother transient evolution than its
Quintom$-$ counterpart. The left panel compares the normalized average
volume function $\eta^{3}/\eta_{0}^{3}$ with the normalized scalar
fields $\psi_{1}/\psi_{10}$ and $\psi_{2}/\psi_{20}$, while the right
panel presents the corresponding deceleration parameter $q(t)$. The
parameters are chosen as $\lambda_{1}=2$, $\lambda_{2}=4$,
$\ell=9/16$, with $r_{1}=r_{2}=2$, $q_{1}=q_{2}=1$, and $n=0.1$,
whereas the remaining constants are determined by the exact analytical
solution. Unlike the Quintom$-$ branch, the accelerated expansion
approaches the de Sitter limit from the quintessence side without
entering a super-accelerated phase. This behavior further illustrates
that different transient dynamics within the mixed kinetic sector
ultimately converge toward the same universal late-time attractor.
}
\label{volumen-quintom-mas}
\end{center}
\end{figure}

In contrast to the Quintom$-$ solution, the Quintom$+$ branch remains
entirely within the accelerated regime without entering a
super-accelerated phase. As shown in Fig.~\ref{volumen-quintom-mas},
the scalar fields evolve monotonically, with $\psi_1$ providing the
dominant contribution at late times, while the deceleration parameter
rapidly converges toward $q=-1$. Therefore, although both quintom
branches originate from the same mixed kinetic structure, they exhibit
qualitatively different transient evolutions before reaching the common
de Sitter attractor.

\newpage

The previous figures illustrate that the six exact branches share a
common asymptotic fate while differing substantially during their
transient evolution. Depending on the kinetic sector and the momentum
constraint, the cosmological solutions exhibit distinct behaviors in
the average expansion rate and in the evolution of the effective
equation-of-state parameter. Quintessence branches remain entirely
within the accelerated regime, phantom branches display either a smooth
accelerated evolution or a transition from deceleration to
acceleration, whereas the mixed (quintom) configurations interpolate
between both behaviors and include a transient super-accelerated phase
for the $m_{12}^{-}$ solution. Despite these differences, every
physically admissible branch evolves toward the same asymptotic de
Sitter attractor. For convenience, the main qualitative properties of
the six exact cosmological solutions are summarized in
Table~\ref{tab:summary_branches}.

\begin{table*}[t]
\centering
\caption{
Qualitative comparison of the six exact cosmological branches obtained
in the generalized Saez--Ballester--K-essence-like theory. Although all
solutions asymptotically converge toward the same de Sitter attractor
($w_{\rm eff}\rightarrow-1$), their transient evolution depends on the
underlying scalar-field kinetic sector. The table summarizes the main
physical characteristics of each exact branch, see Figure \ref{fig:q_comparison_six_branches}.
}
\label{tab:summary_branches}

\begin{tabular}{lccc}
\hline
Model & Branch & Transient behavior of $w_{\rm eff}$ & Characteristic evolution \\
\hline

Phantom
&
$m_{12}^{+}$
&
$-1<w_{\rm eff}<-\frac13$
&
Regular accelerated expansion \\

Phantom
&
$m_{12}^{-}$
&
$w_{\rm eff}>-\frac13
\;\rightarrow\;
-1<w_{\rm eff}<-\frac13$
&
Transition from decelerated to accelerated expansion \\

Quintessence
&
$m_{12}^{+}$
&
$-1<w_{\rm eff}<-\frac13$
&
Accelerated expansion with distinct scalar-field growth \\

Quintessence
&
$m_{12}^{-}$
&
$w_{\rm eff}\simeq-1$
throughout the evolution
&
Near de Sitter evolution \\

Quintom
&
$m_{12}^{+}$
&
$-1<w_{\rm eff}<-\frac13$
&
Mixed kinetic evolution approaching de Sitter \\

Quintom
&
$m_{12}^{-}$
&
$w_{\rm eff}<-1$
&
Transient super-accelerated phase \\

\hline
\end{tabular}
\end{table*}

\section{Directional anisotropy and average observables}
\label{subsec:directional_anisotropy}

The Bianchi type I character of the model is encoded in the directional
scale factors. In the parametrization used in this work, the line
element can be written as \eqref{bianchi}
\begin{equation}
ds^2=-N^2dt^2+\eta^2(t)
\left[
m_1^2(t)dx^2+m_2^2(t)dy^2+m_3^2(t)dz^2
\right],
\end{equation}
where $\eta(t)$ describes the average expansion, while the functions
$m_i(t)$ contain the anisotropic degrees of freedom. The directional
scale factors are
\begin{equation}
a_i(t)=\eta(t)m_i(t),
\end{equation}
and the average comoving volume is
\begin{equation}
V(t)=a_1(t)a_2(t)a_3(t)=\eta^3(t).
\end{equation}

The anisotropic functions satisfy the constraint
\begin{equation}
m_1(t)m_2(t)m_3(t)=1,
\label{mi_constraint}
\end{equation}
which guarantees that the directional functions do not change the
average volume, but only redistribute the expansion among the three
spatial directions. In terms of the Hamiltonian variables, one has
\begin{equation}
u_i(t)=u_{i0}-12n_i\Delta t,
\end{equation}
and 
\begin{equation}
m_i(t)=m_{i0}\exp[-12n_i\Delta t],
\qquad
m_{i0}=e^{u_{i0}}.
\end{equation}
The constants must satisfy
\begin{equation}
m_{10}m_{20}m_{30}=1,
\qquad
n_1+n_2+n_3=0,
\end{equation}
so that Eq.~\eqref{mi_constraint} is preserved at all times.

The isotropic FLRW limit is recovered only when
\begin{equation}
m_1(t)=m_2(t)=m_3(t)=1,
\end{equation}
or equivalently when the anisotropic integration constants vanish.
Solutions with $m_1(t)\neq m_2(t)\neq m_3(t)$ represent genuinely anisotropic Bianchi type I geometries, even though the cosmological observables reconstructed below are written in terms of the average scale factor $\eta(t)$.

\begin{figure}[h]
    \centering
    \includegraphics[width=0.4\textwidth]{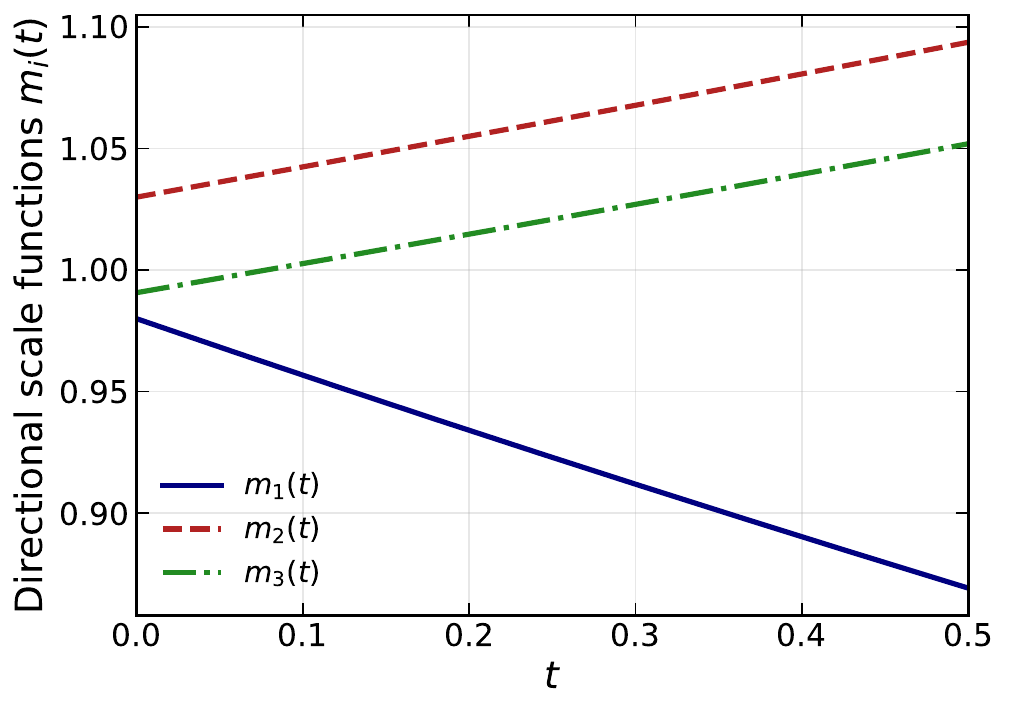}
    \caption{
Representative evolution of the directional scale functions
$m_i(t)$ obtained from the exact anisotropic solution. The three
functions describe the expansion along the independent spatial
directions and satisfy the constraint
$m_1(t)m_2(t)m_3(t)=1$, ensuring that the anisotropic contribution does
not modify the average volume expansion. This example illustrates how
the anisotropic degrees of freedom are entirely encoded in the
directional functions $m_i(t)$, whereas the average cosmological
evolution is determined by $\eta(t)$.
}
    \label{ani1}
\end{figure}

\begin{figure}[t]
    \centering
    \includegraphics[width=0.4\textwidth]{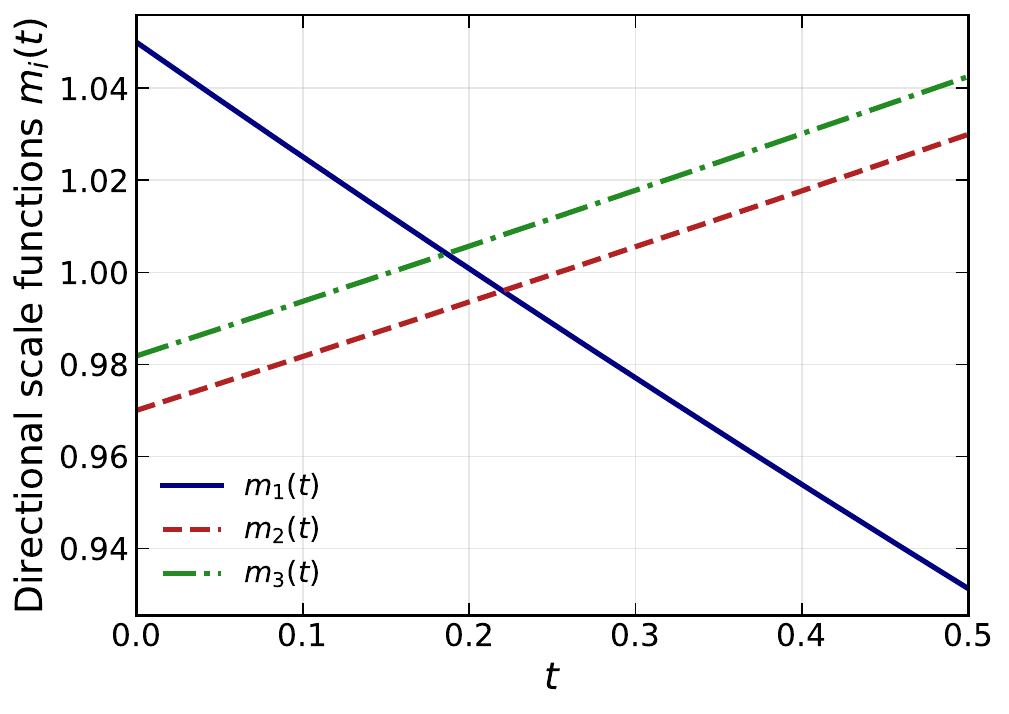}
 \caption{
Representative evolution of the directional scale functions
$m_i(t)$ for a different choice of anisotropic integration constants.
Compared with Fig.~\ref{ani1}, the directional expansion follows a
different anisotropic pattern while preserving the constraint
$m_1(t)m_2(t)m_3(t)=1$. This demonstrates that different anisotropic
configurations may share the same average expansion history described
by $\eta(t)$ but differ in their directional evolution.
}
    \label{ani2}
\end{figure}

Figures~\ref{ani1} and \ref{ani2} show two
representative choices of the anisotropic integration constants
satisfying the Bianchi type I constraints. These examples illustrate
how the directional functions can evolve differently while preserving
the average volume condition $m_1m_2m_3=1$.

The Hubble parameter, the deceleration parameter, and the effective equation-of-state parameter reconstructed in the following subsections are average quantities associated with $\eta(t)$.
:
\begin{equation}
H(t)=\frac{\dot{\eta}}{\eta},
\qquad
q(t)=-\frac{\eta\ddot{\eta}}{\dot{\eta}^{\,2}},
\qquad
w_{\rm eff}(t)=-1-\frac{2}{3}\frac{\dot H}{H^2}.
\end{equation}
They characterize the mean expansion history of the anisotropic
Bianchi type I universe, not an isotropic FLRW geometry. The latter is
obtained only after imposing the additional condition
$m_i(t)=1$ for all spatial directions.

\section{ Comparative analysis}
\subsection{Cosmological evolution of the Hubble parameter}

An additional characterization of the exact cosmological solutions is
provided by the Hubble parameter, which measures the instantaneous
expansion rate of the universe. Since the average scale factor is
identified as

\begin{equation}
a(t)=\eta(t),
\end{equation}

the Hubble parameter is defined in the usual way as

\begin{equation}
H(t)=\frac{\dot a}{a}
     =\frac{\dot\eta}{\eta}.
\label{Hubble1}
\end{equation}

Because the exact solutions are obtained in terms of the average volume
function,

\begin{equation}
V(t)=a^3(t)=\eta^3(t),
\end{equation}

the Hubble parameter can be expressed directly as

\begin{equation}
H(t)
=\frac{1}{3}
\frac{d}{dt}
\ln\!\left[\eta^3(t)\right].
\label{Hubble2}
\end{equation}

The exact analytical expression for $\eta^3(t)$ immediately determines the corresponding Hubble parameter, without requiring additional assumptions or numerical integrations.

To facilitate the comparison among the different exact solutions, the
Hubble parameter is normalized with respect to its present value,

\begin{equation}
E(t)\equiv\frac{H(t)}{H_0},
\end{equation}

where

\begin{equation}
H_0\equiv H(t_0)
\end{equation}

denotes the Hubble parameter evaluated at the present cosmic time.
Throughout this work, the dimensionless instant

\[
t_0=4
\]

is adopted as the present epoch. The normalization condition ensures that $\frac{H(t_0)}{H_0}=1$ for every solution. Accordingly, the value $H/H_0=1$ shown in the figure represents the current expansion rate, whereas deviations from unity describe how the expansion history departs from the present cosmological state.

\begin{figure}[h]
    \centering
    \includegraphics[width=0.4
    \textwidth]{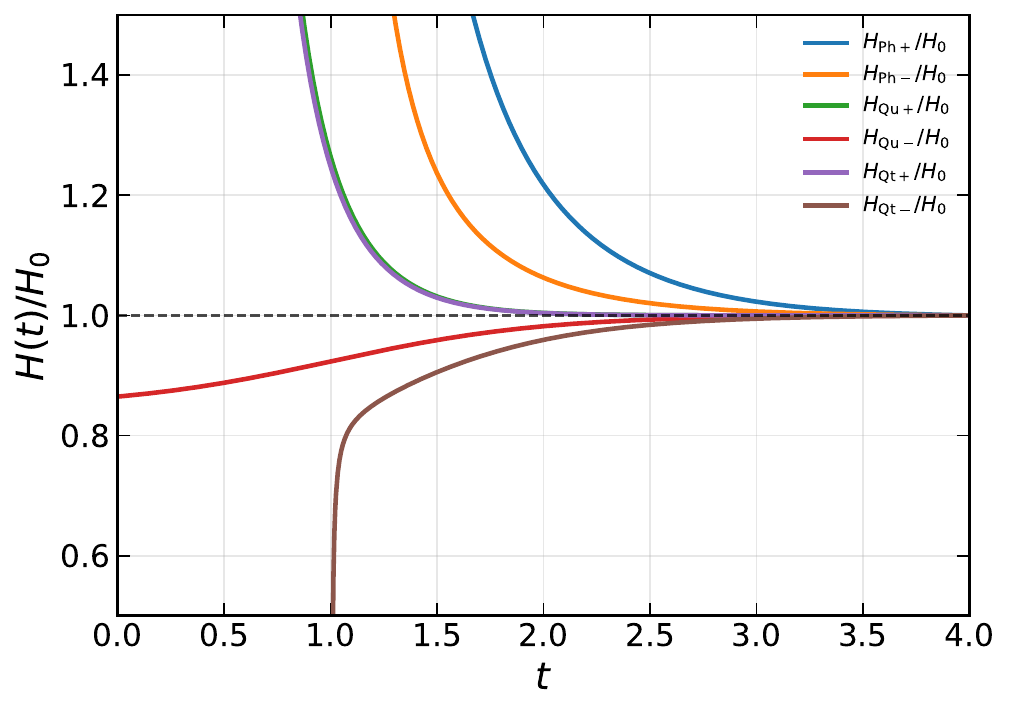}
    \caption{
    Evolution of the normalized Hubble parameter,
    $H(t)/H_0$, corresponding to the six exact cosmological solutions
    analyzed in this work. The normalization is performed with respect
    to the present value $H_0=H(t_0)$, adopting the dimensionless time
    $t_0=4$ as the present epoch. With this normalization, all curves satisfy $H/H_0=1$ at $t=4$, allowing a direct comparison of their relative expansion histories. Although all branches converge to the same normalized present value, they exhibit markedly different transient behaviors before reaching the current cosmological state.
    }
    \label{fig:Hubble}
\end{figure}

Figure~\ref{fig:Hubble} compares the normalized Hubble parameter for the
phantom, quintessence, and quintom branches. Since all solutions are
normalized using their present values, they intersect at
$H/H_0=1$ when $t=4$. This common point is a consequence of the adopted
normalization and should not be interpreted as indicating identical
cosmological dynamics.
The physical information is contained in the different trajectories
followed before reaching the present epoch. The blue, orange, green, and
purple branches begin with expansion rates larger than their current
values, subsequently decreasing monotonically until approaching the
present expansion rate. This behavior indicates that the universe
undergoes a gradual relaxation from an initially faster expansion toward
its present state. In contrast, the red and brown branches start with
$H/H_0<1$, implying that the expansion rate increases with cosmic time
before asymptotically reaching its present value. The sharp variation
displayed by the brown branch near the beginning of the physical
interval originates from the singular structure of the corresponding
exact solution and reflects the lower limit of its physically admissible
domain.

An important result emerging from this analysis is that all exact
solutions converge toward a constant Hubble parameter at late times,
which is fully consistent with the behavior of the deceleration
parameter obtained previously, namely

\[
q(t)\rightarrow-1.
\]

Although the different branches exhibit distinct transient histories, they all evolve toward an asymptotic de Sitter regime characterized by an approximately constant expansion rate. Their principal distinction lies not in the final state, but in the manner in which they approach it.

Some branches
experience a progressive decrease of the expansion rate, whereas others
display an increasing expansion history before relaxing toward the same
late-time attractor. The normalized Hubble parameter serves as a complementary cosmological observable, clearly distinguishing the expansion histories of the different exact solutions while confirming their common asymptotic de Sitter behavior.
\subsection{Comparison of the deceleration parameter}
\label{subsec:q_comparison}

The previous sections analyzed the deceleration parameter for each exact
cosmological branch separately. Here we compare the six branches in a
unified framework in order to identify both the common features of the
model and the qualitative differences in their transient evolution.

The deceleration parameter is reconstructed from the average scale
factor $\eta(t)$ according to
\begin{equation}
q(t)=
-\frac{\eta(t)\ddot{\eta}(t)}
{\dot{\eta}^{\,2}(t)}.
\end{equation}
Since $\eta(t)$ corresponds to the average scale factor of the Bianchi
type I spacetime, the resulting quantity should be interpreted as an
effective deceleration parameter describing the mean cosmological
expansion. The anisotropic degrees of freedom remain encoded in the
directional scale functions $m_i(t)$ and therefore are not represented
directly by $q(t)$.

\begin{figure}[h!]
    \centering
    \includegraphics[width=0.48\textwidth]{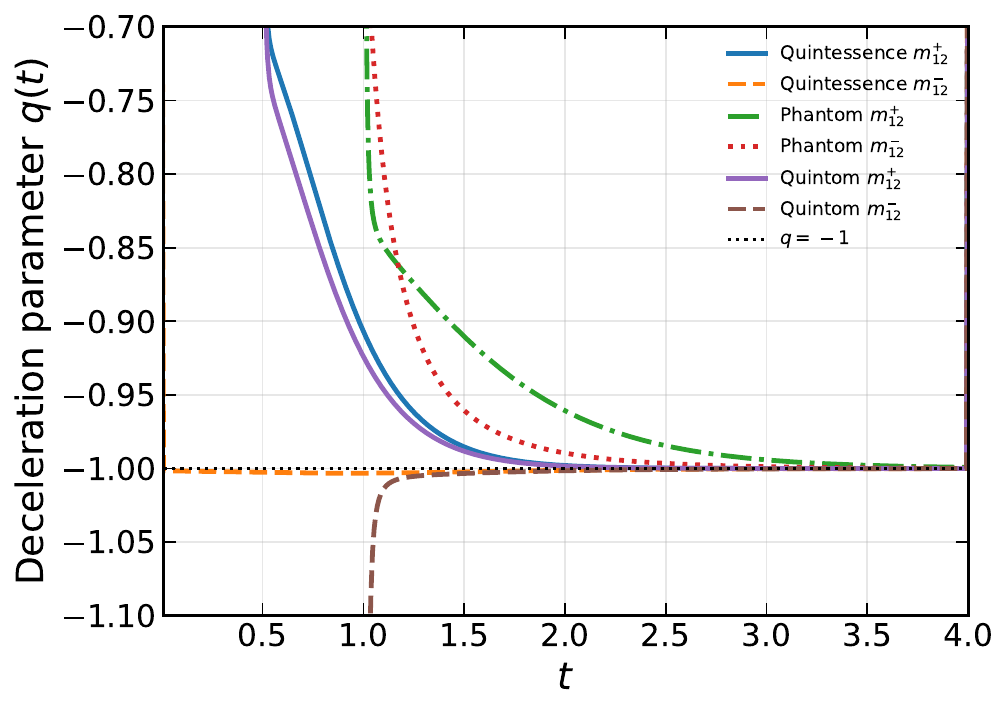}
    \caption{
Comparison of the effective deceleration parameter $q(t)$ for the six
exact cosmological branches obtained in the generalized
Saez--Ballester--K-essence-like Bianchi type I model. The horizontal
dotted line denotes the de Sitter value $q=-1$. Although the different
kinetic sectors produce distinct transient behaviors, all branches
converge toward the same asymptotic attractor, demonstrating the
universality of the late-time average expansion. Since $q(t)$ is
reconstructed from the average scale factor $\eta(t)$, it characterizes
the mean expansion of the anisotropic spacetime, while the directional
anisotropies remain encoded in the functions $m_i(t)$.
}
    \label{fig:q_comparison_six_branches}
\end{figure}

Figure~\ref{fig:q_comparison_six_branches} summarizes the evolution of
$q(t)$ for all six exact cosmological branches. Despite their different
kinetic realizations and parameter constraints, every solution
approaches the asymptotic de Sitter value
\[
q=-1,
\]
demonstrating that the model possesses a universal late-time attractor.
The physical differences among the branches are therefore confined to
their transient evolution. In particular, the quintessence and most
phantom branches approach the de Sitter limit from the quintessence
side, whereas the Quintom$-$ branch temporarily evolves through a
super-accelerated regime ($q<-1$) before converging to the same
asymptotic state. This comparison clearly shows that the kinetic sector
primarily determines the intermediate cosmological evolution, while the
late-time dynamics become universal.

\subsection{Effective equation-of-state parameter}

A complementary characterization of the cosmological dynamics is
provided by the effective equation-of-state parameter, which
establishes a direct connection between the expansion history and the
effective cosmic fluid driving the evolution. Once the deceleration
parameter has been reconstructed from the exact solutions, the
effective equation of state follows from

\begin{equation}
w_{\rm eff}(t)=\frac{2q(t)-1}{3}.
\label{weff}
\end{equation}

This quantity provides a convenient classification of the cosmological
evolution. In particular,
$w_{\rm eff}>-1/3$ corresponds to decelerated expansion,
$-1<w_{\rm eff}<-1/3$ describes accelerated expansion driven by a
quintessence-like effective fluid,
$w_{\rm eff}=-1$ represents the de Sitter limit, whereas
$w_{\rm eff}<-1$ characterizes a phantom-like super-accelerated
regime.

\begin{figure}[t]
\centering
\includegraphics[width=0.5\textwidth]{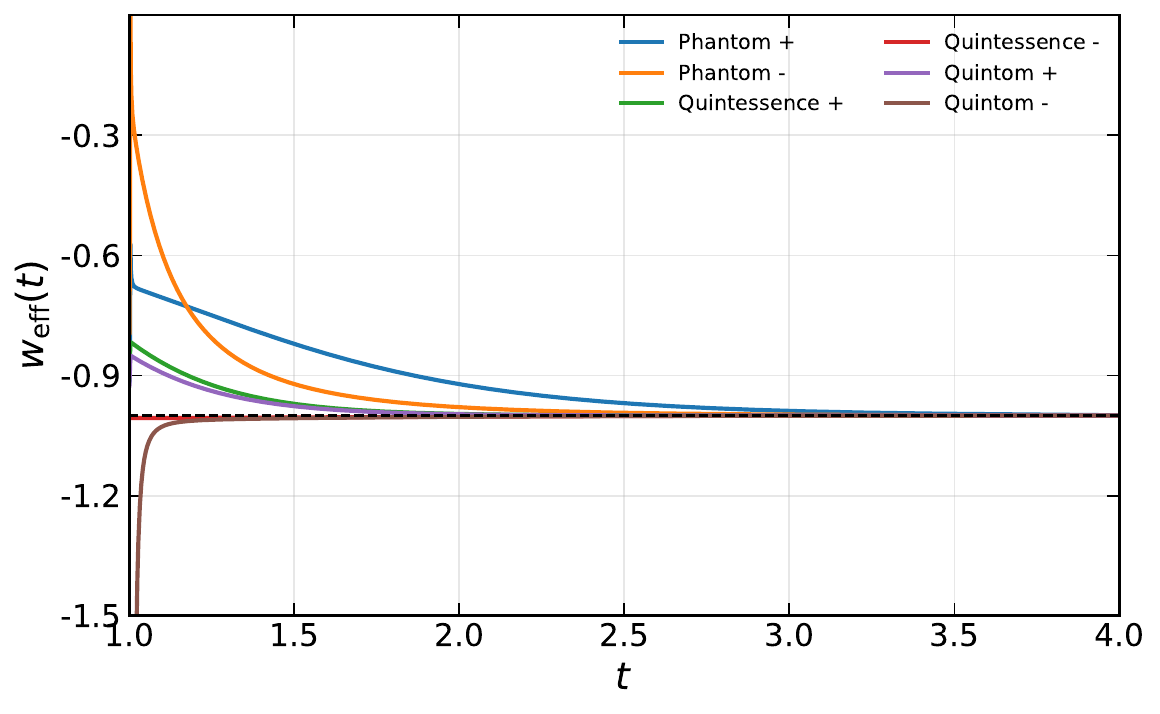}
\caption{
Evolution of the effective equation-of-state parameter
$w_{\rm eff}(t)$ for the six exact cosmological branches.
The horizontal dashed line denotes the de Sitter value,
$w_{\rm eff}=-1$. The Phantom$+$, Phantom$-$,
Quintessence$+$, and Quintom$+$ branches evolve within the accelerated
quintessence regime,
$-1<w_{\rm eff}<-1/3$, approaching the cosmological-constant limit
from above. In contrast, the Quintom$-$ branch temporarily enters a
phantom regime with $w_{\rm eff}<-1$ before converging toward the same
asymptotic state, whereas the Quintessence$-$ solution remains very
close to $w_{\rm eff}=-1$ throughout the evolution. Despite their
different transient behaviors, all six branches converge toward a
common late-time de Sitter attractor.
}
\label{fig:weff}
\end{figure}

Figure~\ref{fig:weff} summarizes the evolution of the effective
equation-of-state parameter for the six exact cosmological branches.
Although the solutions originate from different kinetic sectors, they
share a common asymptotic behavior,
\[
w_{\rm eff}\rightarrow-1,
\]
demonstrating once again the existence of a universal late-time de
Sitter attractor. This result is fully consistent with the evolution of
the Hubble and deceleration parameters discussed in the previous
subsections.

The main physical differences among the branches are therefore encoded
in their transient evolution. The Phantom$+$, Phantom$-$,
Quintessence$+$, and Quintom$+$ branches remain entirely within the
accelerated quintessence regime,
\[
-1<w_{\rm eff}<-\frac13,
\]
while the Quintessence$-$ solution stays remarkably close to the de
Sitter limit throughout the whole evolution. In contrast, the
Quintom$-$ branch is the only solution that temporarily enters the
phantom regime,
\[
w_{\rm eff}<-1,
\]
before smoothly converging toward the common attractor.

An important implication of these results is that the classification of
a solution as phantom, quintessence, or quintom refers to the kinetic
structure of the underlying scalar-field theory rather than to the
effective cosmological equation of state itself. Different scalar-field
realizations produce distinct transient expansion histories, whereas
their late-time average dynamics become practically indistinguishable.
Together with the average volume, the scalar fields, the Hubble
parameter, and the deceleration parameter, the effective
equation-of-state parameter completes a coherent physical description
of the six exact cosmological branches.

\subsection{Unified physical interpretation}

The cosmological observables reconstructed in the previous subsections
provide a coherent physical interpretation of the six exact solutions.
Although these branches originate from different Hamiltonian
structures and distinct scalar-field realizations, they exhibit a
remarkable common feature: all of them evolve toward the same
late-time de Sitter configuration.

The normalized Hubble parameter shows that every branch approaches the
same asymptotic expansion rate, whereas the deceleration parameter
demonstrates that this regime corresponds to accelerated expansion.
The effective equation-of-state parameter further distinguishes the
different branches by identifying the physical nature of their
transient evolution.

More specifically, the Phantom$+$, Quintessence$+$, and Quintom$+$
branches remain entirely within the accelerated quintessence regime,
\[
-1<w_{\rm eff}<-\frac13,
\]
before asymptotically approaching the cosmological-constant limit. The
Phantom$-$ branch evolves from an initially less accelerated phase
toward the same attractor, while the Quintessence$-$ solution remains
very close to the de Sitter limit throughout the entire evolution. In
contrast, the Quintom$-$ branch is the only solution that temporarily
enters a phantom regime,
\[
w_{\rm eff}<-1,
\]
before smoothly converging toward
\[
w_{\rm eff}=-1.
\]

Taken together, these results show that the Hamiltonian structure
primarily controls the intermediate cosmological evolution, whereas the
late-time behavior is remarkably robust against the particular kinetic
realization of the scalar sector. Consequently, the six exact branches
are distinguished not by their final cosmological state, but by the
different evolutionary paths through which they approach a common
de Sitter attractor. The normalized Hubble parameter, the deceleration
parameter, and the effective equation-of-state parameter consistently
capture this universal picture.

\section{Discussion}

The exact analytical solutions derived in this work reveal that the
cosmological evolution of the generalized
S\'aez--Ballester--K-essence-like theory is governed by the interplay
between the kinetic structure of the scalar sector and the common
inverse power-law scalar potential. While the different kinetic
realizations determine the transient cosmological evolution, the
inverse power-law potential drives all physically admissible solutions
toward the same asymptotic accelerated state. Consequently, the
quintessence, phantom, and quintom branches exhibit markedly different
intermediate histories while sharing an identical late-time
cosmological behavior.

One of the most significant results is the emergence of a universal de
Sitter attractor. Although the six exact branches originate from
different Hamiltonian structures and satisfy different parameter
constraints, the reconstructed cosmological observables consistently
approach
\begin{equation}
q\rightarrow-1,\qquad
H/H_0\rightarrow1,\qquad
w_{\rm eff}\rightarrow-1.
\end{equation}
The existence of this common asymptotic behavior does not imply that
the kinetic sector becomes dynamically irrelevant. Instead, the exact
solutions show that the Hamiltonian structure determines the path
followed by the cosmological evolution before the asymptotic regime is
reached. In this sense, the kinetic realization controls the transient
behavior, whereas the late-time average dynamics become largely
independent of the particular scalar-field branch.

The inverse power-law scalar potentials play an essential role in this
behavior. Such potentials have long been studied in connection with
tracker-like cosmological solutions because of their ability to support
a prolonged accelerated expansion. Within the present framework, they
naturally lead the cosmological evolution toward a de Sitter phase even
in the presence of anisotropy and two interacting scalar degrees of
freedom. The exact analytical solutions obtained here therefore show
that this mechanism remains robust beyond isotropic single-field
cosmologies.

Another important aspect concerns the role of the chiral interaction.
Rather than behaving as two independent scalar fields, the chiral
degrees of freedom evolve as a coupled dynamical system in which the
interaction continuously redistributes the effective contribution of
each scalar field during the cosmological evolution. This internal
coupling gives rise to the six exact cosmological branches and explains
why different kinetic realizations produce substantially different
transient expansion histories while preserving the same asymptotic
behavior. The chiral interaction therefore modifies the intermediate
dynamics without altering the universal late-time attractor.

The anisotropic character of the solutions also deserves special
attention. Although the cosmological observables reconstructed in this
work, namely the Hubble parameter, the deceleration parameter, and the
effective equation-of-state parameter, are defined from the average
scale factor, the underlying geometry remains genuinely anisotropic.
The directional scale functions encode the anisotropic degrees of
freedom of the Bianchi type I spacetime, whereas the average scale
factor describes only the mean cosmological expansion. Consequently,
the reconstructed observables should be interpreted as effective
background quantities characterizing the average evolution of an
anisotropic universe rather than observables associated with an
isotropic FLRW spacetime.

More generally, the availability of exact analytical solutions makes it
possible to identify qualitative properties of the theory that are
often difficult to extract from purely numerical analyses. In
particular, the exact integration reveals the existence of six
cosmological branches, clarifies the parameter constraints imposed by
the mixed kinetic interaction, and demonstrates explicitly the
emergence of a universal late-time de Sitter attractor. Together, these
results provide a coherent physical picture of how anisotropy, chiral
interactions, and inverse power-law scalar potentials combine to govern
the cosmological evolution.


\section{Final remarks}

In this work we have investigated exact anisotropic Bianchi type I
cosmological solutions within a generalized
S\'aez--Ballester--K-essence-like theory containing two interacting
scalar fields with quintessence, phantom, and mixed (quintom) kinetic
sectors. Assuming inverse power-law scalar potentials, the field
equations were reduced to integrable forms, allowing the construction
of six exact cosmological branches corresponding to the different
kinetic realizations of the theory.

The exact analytical solutions reveal that all physically admissible
branches share a universal late-time de Sitter attractor despite
displaying substantially different transient cosmological evolutions.
The physical differences among the solutions are therefore encoded in
their intermediate dynamics, whereas their asymptotic behavior is
characterized by
\[
q\rightarrow-1,\qquad
H/H_0\rightarrow1,\qquad
w_{\rm eff}\rightarrow-1.
\]
Among the six branches, the Quintom$-$ solution is the only one that
temporarily enters a phantom regime before converging toward the common
late-time attractor.

An important outcome of the present work is that the exact analytical
solutions permit the explicit reconstruction of physically relevant
cosmological observables. In particular, the normalized Hubble
parameter, the deceleration parameter, and the effective
equation-of-state parameter provide a direct physical interpretation of
the mathematical solutions and consistently demonstrate that the six
exact branches differ only through their transient evolution while
sharing the same asymptotic cosmological state.

The directional scale functions further demonstrate that the exact
solutions remain intrinsically anisotropic throughout the cosmological
evolution. Although the reconstructed observables describe the average
expansion, the anisotropic degrees of freedom continue to be encoded in
the directional functions associated with the Bianchi type I geometry.
Consequently, the present solutions should be regarded as genuine
anisotropic cosmological models rather than effective FLRW
descriptions.

Several extensions of the present framework deserve further
investigation. An immediate step is the reconstruction of observable
quantities such as the redshift dependence of the Hubble parameter,
luminosity distances, and effective energy densities, allowing a direct
comparison with current cosmological observations. Likewise, a complete
analysis of cosmological perturbations and the stability of the exact
solutions will be essential for assessing the physical viability of the
different branches. Extending the formalism to more general
anisotropic geometries or to alternative scalar-field potentials may
also help determine whether the universal late-time de Sitter
attractor identified here is a generic prediction of chiral
multifield cosmological models.

A particularly interesting direction for future research is to extend
the present chiral multifield framework by coupling the scalar sector
to additional matter fields. Such an extension would allow the exact
Bianchi type I solutions derived here to serve as analytical
anisotropic backgrounds for investigating a broad class of
particle-physics motivated cosmological scenarios. Possible
applications include scalar-mediated neutrino interactions,
neutrino-induced dark energy models, and other interacting matter
sectors beyond the standard cosmological framework. Developing these
extensions would require introducing the corresponding interaction
terms into the generalized
S\'aez--Ballester--K-essence-like theory and deriving the modified
field equations. Although this lies beyond the scope of the present
work, the availability of exact anisotropic solutions provides a
promising starting point for exploring the interplay between
fundamental particle interactions, anisotropic cosmological dynamics,
and the origin of dark energy.

The analytical framework developed here demonstrates how exact
solutions can provide physical insight beyond their mathematical
construction. By combining anisotropic cosmology, chiral multifield
dynamics, and exact integrability, the present work establishes a
foundation for exploring more general scalar-field cosmologies and
their possible connections with fundamental physics.

\acknowledgments{ \noindent This work was partially supported by
PROMEP grants UGTO-CA-3. J.S. was partially supported SNI-CONACYT,
A.G. and J.L.P. acknowledge support from SECIHTI post-doctoral
fellowships. Many calculations where done by Symbolic Program REDUCE
3.8}


\begin{thebibliography}{99}
\bibitem{PRD-1999} Jean-Philippe Uzan, Phys. Rev. D {\bf 59}, 123510 (1999), {\it Cosmological scaling solutions of nonminimally coupled
    scalar fields}.
\bibitem{EPJC-2014} Tiberiu Harko, Francisco S.N. Lobo, M.K. Mak, Eur. Phys. J. C  {\bf 74}, 2784 (2014), {\it Arbitrary scalar-field and quintessence
    cosmological models}
\bibitem{ASS-2014} Murli Manohar Verma and Shankar Dayal Pathak, Astrophysics Space Sci. {\bf 350}, 381 (2014),
    {\it Cosmic expansion driven by real scalar field for different forms of potential}


\bibitem{CQG-2014} C.R Fadragas, G. Leon and  Emmanuel N Saridakis, Classical
and Quantum Grav {\bf 31}, 075018  (2014),
    {\it Dynamical analysis of anisotropic scalar-field cosmologies for a wide range of potentials }
\bibitem{CQG-2020P} A. Paliathanasis, Class. Quant. Grav. {\bf 37}, 19 (2020), {\it Dynamics of a chiral cosmology}.
\bibitem{PRD-2017} J. Dutta, W. Khyllep and  N. Tamanini, Phys. Rev. D {\bf 95}, 023515 (2017),
    {\it Scalar-Fluid interacting dark energy: cosmological dynamics beyond the exponential potential}, [arXiv:1701.00744].

\bibitem{PDU-2024} Genly Leon, Alan Coley, Andronikos Paliathanasis, Jonathan Tot and Balkar Yildirim, Physics of the Dark Universe {\bf 45}, 101503 (2024),
    {\it Global dynamics of two models for Quintom Friedman-Lemaitre-Robertson-Walker universes}.

\bibitem{IJTP-2014} Abraham Espinoza Garc\'ia,  J. Socorro and Luis O. Pimentel, Int. J. of Theor. Phys. {\bf 53} (9), 3066-3077 (2014),
    {\it Quantum Bianchi type IX cosmology in K-essence theory.} DOI: 10.1007/s10773-014-2102-0 

\bibitem{roland} Roland de Putter and Eric V. Linder, \emph{Astropart. Phys.} {\bf 28}, 263 (2007).
     {\it Kinetic k-essence and Quintessence}. [arXiv:0705.0400].

\bibitem{chiba} T. Chiba, S. Dutta and R.J. Scherrer, \emph{Phys. Rev. D} {\bf 80}, 043517 (2009).
     {\it Slow-roll k-essence}, [arXiv:0906.0628].

\bibitem{bose} N. Bose and A.S. Majumdar, \emph{Phys. Rev. D} {\bf 79}, 103517 (2009).
     {\it A k-essence model of inflation, dark matter and dark energy}, [arXiv:0812.4131].

\bibitem{arroja} F. Arroja and M. Sasaki, \emph{Phys. Rev. D} {\bf 81}, 107301 (2010).
     {\it A note on the equivalence of a barotropic perfect fluid with a k-essence scalar field}, [arXiv:1002.1376].

\bibitem{tejeiro} L.A. Garc\'ia, J.M. Tejeiro and L. Casta\~neda, {\it K-essence scalar field as dynamical dark energy},
    [arXiv:1210.5259].
\bibitem{AHEP-2014}   J. Socorro, Luis O. Pimentel and Abraham Espinoza Garc\'ia, Advances in High Energy Phys. 805164  (2014),
     {\it Classical Bianchi type I  cosmology  in  K-essence
     theory}

\bibitem{universe-2023} J. Socorro and J. Juan Rosales, Universe {\bf 9}, 185 (2023),  {\it Quantum fractionary cosmology: K-essence theory}
\bibitem{Fractal-2023} J. Socorro, J. Juan Rosales and L. Toledo-Sesma, Fractal Fract.  {\bf 7}, 814 (2023), {\it Anisotropic fractional cosmology: K-essence theory}.
\bibitem{universe-2024} J. Socorro, J. Juan Rosales and Leonel Toledo-Sesma, Universe {\bf 10}, 192 (2024), {\it Non commutative classical and Quantum fractionary Cosmology: FRW case.}
\bibitem{GRG-2025} J. Socorro, J. Juan Rosales and Leonel Toledo-Sesma, General Relativity and Gravitation {\bf 57}, 23 (2025), {\it Noncommutative classical
    and Quantum fractionary Cosmology: Anisotropic Bianchi Type I case.}
\bibitem{saez-ballester-1986} D. Saez and V.J. Ballester, Phys. Lett. A {\bf 113}, 467 (1986).
\bibitem{fizika-2010} M. Sabido, J. Socorro and L. Arturo Ure\~na-L\'opez,
Fizika B, {\bf 19} (4), 177-186 (2010),
    {\it Classical and quantum Cosmology of the S\'aez-Ballester theory},
    [arXiv:0904.0422]
\bibitem{rmf-2010}  J. Socorro, M. Sabido, M.A. S\'anchez G. and M.G. Fr\'ias
Palos, Rev. Mex. F\'is. {\bf 56}(2), 166-171
    (2010), {\it Anisotropic cosmology in S\'aez-Ballester theory: classical and quantum solutions},
     [arxiv:1007.3306].
\bibitem{GC-2012} R.R. Abbyazov and S.V. Chervon, Grav. Cosmol. {\bf 18}, 262 (2012)
\bibitem{QM-2013} S.V. Chervon, Quantum matter {\bf 2}, 71 (2013).
\bibitem{coupled} N. Dimakis and A. Paliathanasis, Class. Quant. Grav. {\bf 38}, (2021), {\it Crossing the phantom divide line
    as a effect of quantum transistions}

\bibitem{PRD-2004-chimento} Luis P. Chimento, Phys. Rev. D {\bf 69}, 123517 (2004), {Extended tachyon field, Chaplygin gas, and solvable
        k-essence cosmologies}.
\bibitem{CQG-2021} J. Socorro, S. P\'erez-Pay\'an, Rafael Hern\'andez-Jim\'enez, Abraham Espinoza-Garc\'ia and Luis Rey D\'iaz-Barr\'on,
    Class. Quantum Grav. {\bf 38}, 135027 (2021), {\it Classical and quantum exact solutions for a FRW  in chiral like cosmology.} [arXiv:2012.11108,
    (gr-qc)]
\bibitem{IJMPD-2021}  Luis Rey D\'iaz-Barr\'on, S. P\'erez-Pay\'an, Abraham Espinoza-Garc\'ia  and J. Socorro, Int. J. Mod. Phys. D {\bf 30} (11), 2150080 (2021),
    {\it Anisotropic chiral cosmology: exact solutions}

\bibitem{Universe-2022P} A. Paliathanasis, Universe {\bf 8}, 199 (2022), {\it Hyperbolic inflation in the Jordan frame}
\bibitem{Tripathy2012} S.K. Tripathy, Astrophys. Sp. Sci. {\bf 340}, (2012) 211.
\bibitem{fr} J. Socorro, Juan Luis P\'erez, Luis Rey D\'iaz-Barr\'on, Abraham Espinoza Garc\'ia, Sinuh\'e P\'erez Pay\'an,  {\it F(R,..) theories
    from the point of view of the Hamiltonian approach: non-vacuum Anisotropic Bianchi type I cosmological
    model}, [arXiv:2512.10850 [gr-qc]]
\end{thebibliography}
\end{document}